\documentclass[epsfig,12pt]{article}
\usepackage{epsfig}
\usepackage{graphicx}
\usepackage{hyperref}

\usepackage{array}
\usepackage{amsmath}
\usepackage{amssymb}

\newcommand{\beq}{\begin{equation}}   
\newcommand{\eeq}{\end{equation}}
\newcommand{\beqn}{\begin{eqnarray}}   
\newcommand{\eeqn}{\end{eqnarray}}

\begin{document}
\unitlength = 1mm

\def\de{\partial}
\def\Tr{ \hbox{\rm Tr}}
\def\const{\hbox {\rm const.}}  
\def\o{\over}
\def\im{\hbox{\rm Im}}
\def\re{\hbox{\rm Re}}
\def\bra{\langle}\def\ket{\rangle}
\def\Arg{\hbox {\rm Arg}}
\def\Re{\hbox {\rm Re}}
\def\Im{\hbox {\rm Im}}
\def\diag{\hbox{\rm diag}}


\def\QATOPD#1#2#3#4{{#3 \atopwithdelims#1#2 #4}}
\def\stackunder#1#2{\mathrel{\mathop{#2}\limits_{#1}}}
\def\stackreb#1#2{\mathrel{\mathop{#2}\limits_{#1}}}
\def\Tr{{\rm Tr}}
\def\res{{\rm res}}
\def\Bf#1{\mbox{\boldmath $#1$}}
\def\balpha{{\Bf\alpha}}
\def\bbeta{{\Bf\beta}}
\def\bgamma{{\Bf\gamma}}
\def\bnu{{\Bf\nu}}
\def\bmu{{\Bf\mu}}
\def\bphi{{\Bf\phi}}
\def\bPhi{{\Bf\Phi}}
\def\bomega{{\Bf\omega}}
\def\blambda{{\Bf\lambda}}
\def\brho{{\Bf\rho}}
\def\bsigma{{\bfit\sigma}}
\def\bxi{{\Bf\xi}}
\def\bbeta{{\Bf\eta}}
\def\d{\partial}
\def\der#1#2{\frac{\d{#1}}{\d{#2}}}
\def\Im{{\rm Im}}
\def\Re{{\rm Re}}
\def\rank{{\rm rank}}
\def\diag{{\rm diag}}
\def\2{{1\over 2}}
\def\ntwo{${\mathcal N}=2\;$}
\def\nfour{${\mathcal N}=4\;$}
\def\none{${\mathcal N}=1\;$}
\def\ntwot{${\mathcal N}=(2,2)\;$}
\def\ntwoo{${\mathcal N}=(0,2)\;$}
\def\x{\stackrel{\otimes}{,}}

\newcommand{\cpn}{CP$(N-1)\;$}
\newcommand{\wcpn}{wCP$_{N,\widetilde{N}}(N_f-1)\;$}
\newcommand{\wcpd}{wCP$_{\widetilde{N},N}(N_f-1)\;$}
\newcommand{\vp}{\varphi}
\newcommand{\pt}{\partial}
\newcommand{\tN}{\widetilde{N}}
\newcommand{\ve}{\varepsilon}
\renewcommand{\theequation}{\thesection.\arabic{equation}}

\newcommand{\sun}{SU$(N)\;$}

\setcounter{footnote}0

\vfill

\begin{titlepage}

\begin{flushright}
FTPI-MINN-18/09, UMN-TH-3718/18\\
October 4, 2018
\end{flushright}

\begin{center}
{  \Large \bf  
 Hadrons of \boldmath{\ntwo} Supersymmetric QCD  in 
\\[2mm]
Four Dimensions  
from  Little String Theory
}

\vspace{5mm}

{\large \bf   M.~Shifman$^{\,a}$ and \bf A.~Yung$^{\,\,a,b,c}$}
\end {center}

\begin{center}

$^a${\it  William I. Fine Theoretical Physics Institute,
University of Minnesota,
Minneapolis, MN 55455}\\
$^{b}${\it National Research Center ``Kurchatov Institute'', 
Petersburg Nuclear Physics Institute, Gatchina, St. Petersburg
188300, Russia}\\
$^{c}${\it  St. Petersburg State University,
 Universitetskaya nab., St. Petersburg 199034, Russia}
\end{center}

\vspace{0.3cm}

\begin{center}
{\large\bf Abstract}
\end{center}

It was  recently shown that non-Abelian vortex strings  supported in a version of four-dimensional 
\ntwo supersymmetric QCD  (SQCD) become critical superstrings. 
In addition to four translational moduli, non-Abelian strings under consideration have six orientational and size moduli. 
Together they form a ten-dimensional target space required for a superstring to 
be critical, namely, the product of the flat four-dimensional
space and conifold -- a non-compact  Calabi-Yau threefold. In this paper we report on further studies
of low-lying closed string states which emerge in four dimensions  and identify them as
hadrons of our four-dimensional  \ntwo SQCD.  We use the approach based on  ``little
string theory,'' describing  critical string  on the conifold as a non-critical $c=1$ string with the Liouville field 
and a compact scalar at 
the self-dual radius. In addition to massless hypermultiplet found earlier we observe several massive vector multiplets
and a massive spin-2 multiplet, all belonging to the long (non-BPS) representations  of \ntwo supersymmetry in four dimensions.
All the above states are interpreted as baryons formed  by a closed string with confined monopoles attached. Our construction presents 
an example of a ``reverse holography."

\vspace{2cm}

\end{titlepage}

\newpage


\newpage

\section {Introduction }
\label{intro}
\setcounter{equation}{0}

In this paper we continue studying the spectrum of four-dimensional ``hadrons" formed by the closed
critical string \cite{SYlittles} which in turn can be obtained from a solitonic vortex string under an appropriate choice of the coupling constant \cite{SYcstring}.
One of our main tasks is to analyze the structure of the 4D supermultiplets emerging from quantization of the closed string mentioned above.
We will start though from a brief review of the setup.

The problem of understanding confining gauge theories splits into two different equally fundamental tasks.
The first one is to understand the physical nature of confinement and describe the formation  of confining strings. 
There was a great progress in this direction in supersymmetric gauge theories  due to the breakthrough papers by
Seiberg and Witten \cite{SW1,SW2} in which the monopole condensation was shown to occur in the monopole vacua
of \ntwo supersymmetric QCD (SQCD). This leads to the formation of Abelian Abrikosov-Nielsen-Olesen (ANO) 
  vortices \cite{ANO} which confine color electric char\-ges. Attempts to find a non-Abelian generalization of this mechanism
led to the discovery of the so called  ``instead-of-confinement'' phase which occurs in the quark vacua of
\ntwo SQCD, see \cite{SYdualrev} for a review. In this phase the (s)quarks condense while the monopoles are confined.

Once  the nature of the confining string is understood the second task is to quantize this string in
four-dimensional (4D) theory outside the critical dimension to study the hadron spectrum. Most solitonic strings,
such as the ANO strings, 
have a finite thickness manifesting itself in the presence of an infinite series of unknown higher-derivative
corrections in the effective sigma model on the string world sheet. This makes the task of quantizing such a 
string virtually impossible.

Recent advances in this direction
 \cite{SYcstring} demonstrated that the non-Abelian solitonic vortex  in a 
particular version of 4D \ntwo SQCD becomes a critical  superstring.
This particular 4D SQCD has the 
 U(2) gauge group, four quark flavors and the Fayet-Iliopoulos (FI) \cite{FI} parameter $\xi$.  

Non-Abelian vortices were first discovered in 
\ntwo SQCD with the U$(N)$ gauge group  and $N_f \ge N$ flavors of quark hypermultiplets
\cite{HT1,ABEKY,SYmon,HT2}.
In addition to four translational moduli characteristic of the  ANO strings 
\cite{ANO}, the non-Abelian strings carry orientational  moduli, as well as the size moduli if $N_f>N$
\cite{HT1,ABEKY,SYmon,HT2} (see \cite{Trev,Jrev,SYrev,Trev2} for reviews). If $N_f>N$ their dynamics
are described by effective two-dimensional sigma model on the string world sheet with 
the target space 
\beq
\mathcal{O}(-1)^{\oplus(N_f-N)}_{\mathbb{CP}^1}\,,
\label{12}
\eeq
to which we will refer to as the weighted CP  model (WCP$(N,N_f-N)$). 

 For $N_f=2N$
the model becomes conformal. Moreover, for $N=2$ the 
dimension of the orientational/size moduli space is six and they can be combined with 
four translational moduli to form a ten-dimensional space required for 
superstring to become critical.\footnote{The non-Abelian vortex string is 1/2
BPS saturated and, therefore,  has \ntwot supersymmetry on its world sheet. Thus, we actually deal with a superstring in the case at hand.}

In this case the target space of the world sheet 2D theory on 
the non-Abelian vortex string is
 $\mathbb{R}^4\times Y_6$, where $Y_6$ is a non-compact six-dimensional Calabi-Yau manifold, the so-called
resolved conifold \cite{Candel,NVafa}. 

Since non-Abelian vortex string on the conifold is critical it has 
a perfectly good UV behavior. This opens the possibility that it can become thin in a certain regime 
\cite{SYcstring}. The string transverse size is given  by $1/m$, where $m$ is 
a typical mass scale  of the four-dimensional fields forming the string. The string cannot be thin in a {\em weakly
coupled} 4D theory because at weak coupling $m\sim g\sqrt{T}$ and is always small in the units of $\sqrt{T}$
where $T$ is the tension. 
 Here  $g$ is the gauge coupling constant  of the 4D ${\cal N}=2$  QCD and $T$ is  the string tension.

A conjecture was put forward in  \cite{SYcstring}  that  at strong coupling 
in the vicinity of a critical value of $g_c^2\sim 1$ the non-Abelian string on the conifold becomes thin,
and higher-derivative corrections in the action can be ignored. It is expected that
the thin string produces linear Regge trajectories even for small spins \cite{SYcstring}.
 The above  conjecture implies\,\footnote{At $N_f=2N$ the beta function
of the 4D \ntwo QCD is zero, so the gauge coupling $g^2$ does not run. Note, however, that conformal
invariance in the 4D theory is broken by the FI parameter $\xi$ which does not run either.} that $m(g^2) \to \infty$ at  $ g^2\to g_c^2$. 

A version of the string-gauge duality
for 4D SQCD  was proposed \cite{SYcstring}: at weak coupling this 
theory is in the Higgs phase and can be 
described in terms of (s)quarks and Higgsed gauge bosons, while at strong coupling hadrons of this theory 
can be understood as string states formed by the non-Abelian vortex string.

The vortices in the U$(N)$ theories under consideration
are topologically stable and cannot be broken. Therefore
the finite-length strings are closed. Thus, we focus on the 
closed strings. The goal is to  identify the closed string states 
with the hadrons of  4D \ntwo SQCD. 

The first
step of this program, namely,  identifying massless
string states was carried out in \cite{KSYcstring,KSYconifold} using supergravity formalism.
In particular,  a single matter hypermultiplet associated with the deformation 
of the complex structure of the conifold was found as the only 4D {\em massless} mode of the string. 
Other states arising from the massless ten-dimensional graviton are not dynamical
in four dimensions. In particular, the 4D graviton and  unwanted vector multiplet associated with
deformations of the K\"ahler form of the conifold are absent.
This is due to non-compactness of the  Calabi-Yau manifold we deal with and 
non-normalizability of the corresponding  modes over six-dimensional space $Y_6$.

The next step was done in \cite{SYlittles} where a number of massive states of the closed non-Abelian vortex 
string was found.   This step required a change of strategy. The point is that
the coupling constant $1/\beta$ of the world sheet WCP(2,2)  is not small. Moreover $\beta$ tends to zero
once the 4D coupling $g^2$ approaches the critical value $g^2_c$ we are interested in. At $\beta\to 0$ the resolved
conifold develops a conical singularity. The supergravity approximation does not work 
for massive states.\footnote{This is in contradistinction to the massless states. 
For the latter, we can perform
computations at large $\beta$ where the supergravity approximation is valid and 
then extrapolate to strong coupling.
In the sigma-model language massless states corresponds to chiral primary operators. They are protected by
\ntwot world-sheet supersymmetry and their masses are not lifted by quantum corrections.}

To analyze the massive states the  little string theories (LST) approach (see  \cite{Kutasov} for a review)
was used in \cite{SYlittles}.
Namely, we used  the equivalence between the 
critical string  on the conifold and non-critical $c=1$ string which contains the Liouville 
field and a compact scalar at 
the self-dual radius \cite{GivKut,GVafa}. The latter theory (in the mirror Wess-Zumino-Novikov-Witten (WZNW) formulation) can be
analyzed by virtue of algebraic methods.  This leads to identification of towers of massive states with spin zero and spin two \cite{SYlittles}.

In this paper we focus on the 4D multiplet structure of the states found earlier in \cite{KSYconifold,SYlittles}. In addition to the massless BPS hypermultiplet
associated with deformations of the complex structure of the conifold we identify several massive vector multiplets and a massive spin-2 multiplet, all belonging to long non-BPS representations  of \ntwo supersymmetry in four dimensions.
 We interpret all states we found  as baryons formed  by a closed string with 
confined monopoles attached. 

Note, that the relation between LST and certain partly confined 4D gauge theories was also
studied in \cite{Dorey1,Dorey2} using AdS/CFT approach. In particular, massive stringy states
were discussed.

The paper is organized as follows. In Sec.~\ref{worldsheet} we review the description of non-Abelian vortex
as a critical superstring on a conifold and identify massless string state. In Sec. \ref{c=1} we review LST approach in terms of non-critical $c=1$ string and the spectrum of  massive states. In Sec.~\ref{b} we 
introduce 4D supercharges and construct massless BPS hypermultiplet. In Sec.~\ref{m=3/2} we consider the lowest massive string excitations and show that they forms a long vector supermultiplet.  Section~\ref{j=1} deals with the construction of \ntwo spin-2 stringy supermultiplet. In Sec.~\ref{Regge} we discuss linear Regge trajectories,
while Section~\ref{conclusions} summarizes our conclusions. In Appendix A we describe the BRST operator and transitions
between different pictures. In Appendix B we review long \ntwo supermultiplets in 4D.

\section {Non-Abelian vortex string }
\label{worldsheet}
\setcounter{equation}{0} 

\subsection{Four-dimensional \boldmath{${\mathcal N}=2\;$} SQCD}

As was already mentioned non-Abelian vortex-strings were first found in 4D
\ntwo SQCD with the gauge group U$(N)$ and $N_f \ge N$ flavors (i.e. the quark hypermultiplets)
supplemented by the FI $D$ term $\xi$
\cite{HT1,ABEKY,SYmon,HT2}, see for example \cite{SYrev} for a detailed review of this theory.
Here we just mention that at weak coupling, $g^2\ll 1$, this theory is in the Higgs phase in which the scalar
components of the quark multiplets (squarks) develop vacuum expectation values (VEVs). These VEVs breaks 
the U$(N)$ gauge group
Higgsing  all gauge bosons. The Higgsed gauge bosons combine with the screened quarks to form long \ntwo multiplets with mass $m \sim g\sqrt{\xi}$.

 The global flavor SU$(N_f)$ is broken down to the so called color-flavor
locked group. The resulting global symmetry is
\beq
 {\rm SU}(N)_{C+F}\times {\rm SU}(N_f-N)\times {\rm U}(1)_B,
\label{c+f}
\eeq
see \cite{SYrev} for more details. 

The unbroken global U(1)$_B$ factor above is identified with a baryonic symmetry. Note that 
what is usually identified as the baryonic U(1) charge is a part of  our 4D theory  gauge group.
 ``Our" U(1)$_B$
is  an unbroken by squark VEVs combination of two U(1) symmetries:  the first is a subgroup of the flavor 
SU$(N_f)$ and the second is the global U(1) subgroup of U$(N)$ gauge symmetry.

\begin{figure}
\epsfxsize=10cm
\centerline{\epsfbox{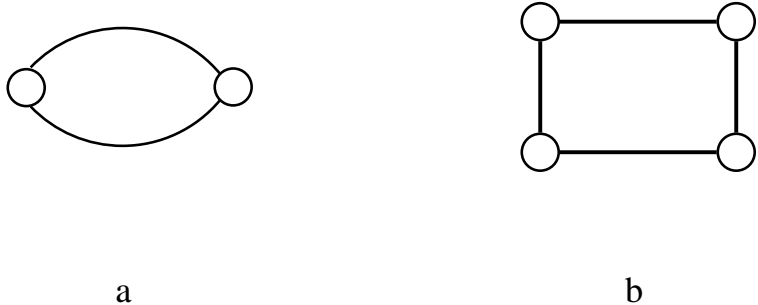}}
\caption{\small  Examples of the monopole ``necklace'' baryons: 
Open circles denote  monopoles. }
\label{baryons}
\end{figure}

As was already noted, we consider \ntwo SQCD  in the Higgs phase:  $N$ squarks  condense. Therefore,  non-Abelian 
vortex strings confine monopoles. In the \ntwo 4D theory these strings are 1/2 BPS-saturated; hence,  their
tension  is determined  exactly by the FI parameter,
\beq
T=2\pi \xi\,.
\label{ten}
\eeq
However, 
the monopoles cannot be attached to the string endpoints. In fact, in the U$(N)$ theories confined  
 monopoles 
are  junctions of two distinct elementary non-Abelian strings \cite{T,SYmon,HT2} (see \cite{SYrev} 
for a review). As a result,
in  four-dimensional \ntwo SQCD we have 
monopole-antimonopole mesons in which the monopole and antimonopole are connected by two confining strings.
 In addition, in the U$(N)$  gauge theory we can have baryons  appearing as  a closed 
``necklace'' configurations of $N\times$(integer) monopoles \cite{SYrev}. For the U(2) gauge group the 
lightest baryon presented by such a ``necklace'' configuration  
consists of two monopoles, see Fig.~\ref{baryons}.

Both stringy monopole-antimonopole mesons and monopole baryons with spins $J\sim 1$ have mass determined by the string tension,  $\sim \sqrt{\xi}$ and are heavier at weak coupling than perturbative states, which have mass 
$m\sim g\sqrt{\xi}$. However, according to our conjecture, at strong coupling near the 
critical point $g^2_c$ 
$m\to\infty$, see \cite{SYcstring} and Sec.~\ref{thinstring} below. In this regime perturbative states decouple
and we are left with hadrons formed by the closed string states.\footnote{There are also massless
bifundamental quarks, charged with respect to both non-Abelian factors in (\ref{c+f}). These are associated with the Higgs branch present in 4D QCD, see \cite{SYrev,KSYconifold} for details.} All hadrons identified as closed string states in this paper turn out to be baryons and look like monopole ``necklaces,'' see 
Fig.~\ref{baryons}.

\subsection{World sheet sigma model}

The presence of color-flavor locked group SU$(N)_{C+F}$ is the reason for the formation of the
non-Abelian vortex strings \cite{HT1,ABEKY,SYmon,HT2} in our 4D SQCD.
The most important feature of these vortices is the presence of the so-called orientational  zero modes.

Let us briefly review the model emerging on the world sheet
of the non-Abelian critical string \cite{SYcstring,KSYcstring,KSYconifold}.
If $N_f=N$  the dynamics of the orientational zero modes of the non-Abelian vortex, which become 
orientational moduli fields 
 on the world sheet, is described by two-dimensional
\ntwot supersymmetric ${\rm CP}(N-1)$ model \cite{SYrev}.

If one adds extra quark flavors, non-Abelian vortices become semilocal.
They acquire size moduli \cite{AchVas}.  
In particular, for the non-Abelian semilocal vortex at hand,  in 
addition to  the orientational zero modes  $n^P$ ($P=1,2$), there are  the so-called size moduli   
$\rho^K$ ($K=1,2$) \cite{AchVas,HT1,HT2,SYsem,Jsem,SVY}.  
The target space of the ${\rm WCP}(2,2)$ sigma model on the string world sheet is defined by the $D$-term condition
\beq
|n^P|^2-|\rho^K|^2 = \beta\,,
\label{Dterm}
\eeq 
and a U(1) phase  is gauged away.  

The total number of real bosonic degrees of freedom in this model is six, 
where we take into account the constraint (\ref{Dterm})
and the fact that one U(1) phase is gauged away. As was already mentioned, 
these six internal degrees of freedom 
are combined with four translational moduli  to form a ten-dimensional space needed 
for superstring to be critical.

At weak coupling the world sheet coupling constant $\beta$  in (\ref{Dterm}) is related to 
the 4D SU(2) gauge coupling as follows:
$g^2$ 
\beq
\beta\approx \frac{4\pi}{g^2}\,,
\label{betag}
\eeq
see \cite{SYrev}.
Note that the first (and the only) coefficient is the same for the 4D SQCD and the
world-sheet model $\beta$ 
functions. Both vanish at $N_f=2N$. This ensures that our world-sheet theory is conformal.

Since non-Abelian  vortex string  is 1/2 BPS it preserves ${\mathcal N} =(2,2)$  in the world sheet
sigma model which 
is necessary to have \ntwo space-time supersymmetry \cite{Gepner,BDFM}. Moreover, as was shown in \cite{KSYconifold},
 the string theory of the non-Abelian critical vortex is type IIA.

The global symmetry of the world-sheet sigma model is
\beq
 {\rm SU}(2)\times {\rm SU}(2)\times {\rm U}(1)\,,
\label{globgroup}
\eeq
i.e. exactly the same as the unbroken global group in the 4D theory, cf. (\ref{c+f}), at  $N=2$ and $N_f=4$. 
The fields $n$ and $\rho$ 
transform in the following representations:
\beq
n:\quad (\textbf{2},\,0,\, 0), \qquad \rho:\quad (0,\,\textbf{2},\, 1)\,.
\label{repsnrho}
\eeq

\subsection{Thin string regime}
\label{thinstring}

The coupling constant of 4D SQCD can be complexified
\beq
 \tau \equiv i\frac{4\pi}{g^2} +\frac{\theta_{4D}}{2\pi}\,,
\label{bulkduality}
\eeq 
 where $\theta_{4D}$ is the four-dimensional $\theta$ angle.
Note that SU$(N)$ version of the four-dimensional SQCD at hand 
possesses a strong-weak coupling duality, namely, $\tau\to  -\frac{1}{\tau}\,$ \cite{ArgPlessShapiro,APS}.
This suggests that the self-dual point $g^2 =4\pi$ would be  a natural candidate for a critical value $g^2_c$,
where our non-Abelian vortex string becomes thin.\footnote{We suggested this earlier in \cite{KSYcstring,KSYconifold}.}
However, as was shown recently in \cite{Karasik}, $S$-duality maps our U$(N)$ theory to a theory in which
a different U(1) subgroup of the flavor group is gauged. In particular,  in our U$(N)$ theory
all quark flavors have equal charges with respect to the U(1) subgroup of the U(2) gauge group, while  in the $S$-dual version only one 
flavor is charged with respect to the U(1) gauge group. As a result, the $ S$-dual version supports  a different type of non-Abelian strings \cite{Karasik}. 

This means that $S$-duality is broken in our 4D theory by the choice of the U(1)
subgroup which is gauged.\footnote{We are grateful to E. Gerchkovitz and A. Karasik
for pointing out to us this circumstance.}
  We do not consider $S$-duality and its consequences here.

The two-dimensional coupling constant $\beta$ can be naturally complexified too if we include the two-dimensional
$\theta$ term,
\beq
\beta \to \beta + i\,\frac{\theta_{2D}}{2\pi}\,.
\eeq 

The exact relation between the complexified 4D and 2D couplings is as follows:
\beq
\exp{(-2\pi\beta)} = - h(\tau)[h(\tau) +2],
\label{taubeta}
\eeq
where the function $h(\tau)$ is a special modular function of $\tau$ defined in terms of the
$\theta$-functions, $$h(\tau)=\theta_1^4/(\theta_2^4-\theta_1^4)\,.$$
 This function enters the Seiberg-Witten curve in our 4D theory
 \cite{ArgPlessShapiro,APS}. Equation (\ref{taubeta}) generalizes the quasiclassical relation
\eqref{betag}. It can be derived using 2D-4D correspondence, namely, the match of the BPS spectra of the 4D theory 
at $\xi=0$ and the world-sheet theory on the non-Abelian string \cite{Dorey,DoHoTo,SYmon,HT2}. Details of this
derivation will be presented elsewhere. Note that our result \eqref{taubeta} differs from that
obtained in \cite{Karasik} using the localization technique.

According to the hypothesis formulated in  \cite{SYcstring}, our critical non-Abelian string becomes infinitely thin at  strong coupling at the critical  point $\tau_c$ (or $g^2_c$).  Moreover, in
\cite{KSYconifold} we conjectured  that $\tau_c$ corresponds to $\beta =0$ in the world-sheet theory via relation \eqref{taubeta}.
Thus, we assume that $m \to \infty$ at $\beta=0$, which corresponds to $g^2=g^2_c$ in 4D SQCD.
 
At the    point $\beta=0$  
the non-Abelian string 
becomes infinitely thin,  higher derivative terms can be neglected and the theory of the non-Abelian 
string reduces to the WCP(2,2) model. The point $\beta=0$ is a natural choice because at this point
 we have a regime change in the 2D sigma model. 
This is the point where the resolved conifold defined by the $D$ term constraint
(\ref{Dterm}) develops a conical singularity \cite{NVafa}.

\subsection {Massless 4D baryon as deformation of the conifold complex structure}
\label{conifold}

In this section we will briefly review the only 4D massless state associated 
with the deformation of the conifold complex structure. It was found in \cite{KSYconifold}.
 As was already mentioned, all other modes arising from the massless 10D
graviton have non-normalizable wave functions over the conifold. In particular, the 4D graviton is
absent \cite{KSYconifold}. This result matches our expectations since from the very beginning we started from
\ntwo SQCD in the flat four-dimensional space without gravity.

The target space of the world sheet WCP(2,2) model is defined by the $D$-term condition (\ref{Dterm}).
 We can construct the U(1) gauge-invariant ``mesonic'' variables
\beq
w^{PK}= n^P \rho^K.
\label{w}
\eeq
These variables are subject to the constraint
${\rm det}\, w^{PK} =0$, or
\beq
\sum_{\alpha =1}^{4} w_{\alpha}^2 =0,
\label{coni}
\eeq
where 
$$w^{PK}\equiv \sigma_{\alpha}^{PK}w_{\alpha}\,,$$ 
and the $\sigma$ matrices above
are  $(1,-i\tau^a)$, $a=1,2,3$.
Equation (\ref{coni}) defines the conifold $Y_6$.  
It has the K\"ahler Ricci-flat metric and represents a non-compact
 Calabi-Yau manifold \cite{Candel,W93,NVafa}. It is a cone which can be parametrized 
by the non-compact radial coordinate 
\beq
\widetilde{r}^{\, 2} =\sum_{\alpha =1}^{4} |w_{\alpha}|^2\,
\label{tilder}
\eeq
and five angles, see \cite{Candel}. Its section at fixed $\widetilde{r}$ is $S_2\times S_3$.

At $\beta =0$ the conifold develops a conical singularity, so both $S_2$ and $S_3$  
can shrink to zero.
The conifold singularity can be smoothed out
in two distinct ways: by deforming the K\"ahler form or by  deforming the 
complex structure. The first option is called the resolved conifold and amounts to introducing 
a non-zero $\beta$ in (\ref{Dterm}). This resolution preserves 
the K\"ahler structure and Ricci-flatness of the metric. 
If we put $\rho^K=0$ in (\ref{Dterm}) we get the $CP(1)$ model with the $S_2$ target space
(with the radius $\sqrt{\beta}$).  
The resolved conifold has no normalizable zero modes. 
In particular, 
the modulus $\beta$  which becomes a scalar field in four dimensions
 has non-normalizable wave function over the 
$Y_6$ manifold \cite{KSYconifold}.  

As  explained in \cite{GukVafaWitt,KSYconifold}, non-normalizable 4D modes can be 
interpreted as (frozen) 
coupling constants in the 4D  theory. 
The $\beta$ field is the most straightforward example of this, since the 2D coupling $\beta$ is
 related to the 4D coupling, see Eq. (\ref{taubeta}).

If $\beta=0$ another option exists, namely a deformation 
of the complex structure \cite{NVafa}. 
It   preserves the
K\"ahler  structure and Ricci-flatness  of the conifold and is 
usually referred to as the {\em deformed conifold}. 
It  is defined by deformation of Eq.~(\ref{coni}), namely,   
\beq
\sum_{\alpha =1}^{4} w_{\alpha}^2 = b\,,
\label{deformedconi}
\eeq
where $b$ is a complex number.
Now  the $S_3$ can not shrink to zero, its minimal size is 
determined by
$b$. 

The modulus $b$ becomes a 4D complex scalar field. The  effective action for  this field was calculated in \cite{KSYconifold}
using the explicit metric on the deformed conifold  \cite{Candel,Ohta,KlebStrass},
\beq
S(b) = T\int d^4x |\pt_{\mu} b|^2 \,
\log{\frac{T^2 L^4}{|b|}}\,,
\label{Sb}
\eeq
where $L$ is the  size of $\mathbb{R}^4$ introduced as an infrared regularization of 
logarithmically divergent $b$ field 
norm.\footnote{The infrared regularization
on the conifold $\widetilde{r}_{\rm max}$ translates into the size $L$ of the 4D space 
 because the variables  $\rho$ in \eqref{tilder} have an interpretation of the vortex string sizes,
$\widetilde{r}_{\rm max}\sim TL^2$ .}

We see that the norm of
the $b$ modulus turns out to be  logarithmically divergent in the infrared.
The modes with the logarithmically divergent norm are at the borderline between normalizable 
and non-normalizable modes. Usually
such states are considered as ``localized'' on the string. We follow this rule.  We can
 relate this logarithmic behavior to the marginal stability of the $b$ state, see \cite{KSYconifold}.

The field $b$,  being massless, can develop a VEV. Thus, 
we have a new Higgs branch in 4D \ntwo SQCD which is developed only for the critical value of 
the coupling constant $g^2_c$. 

The logarithmic metric in (\ref{Sb}) in principle can receive both perturbative and 
non-perturbative quantum corrections in $1/\beta$, the sigma model coupling. However, in the
\ntwo  theory the non-renormalization
theorem of \cite{APS} forbids the  dependence of the Higgs branch metric  on the 4D coupling 
constant $g^2$.
Since the 2D coupling $\beta$ is related to $g^2$ we expect that the logarithmic metric in (\ref{Sb})
will stay intact. This expectation is confirmed in \cite{SYlittles}.

 In \cite{KSYconifold} the massless state $b$ was interpreted as a baryon of 4D \ntwo SQCD.
Let us explain this.
 From Eq.~(\ref{deformedconi}) we see that the complex 
parameter $b$ (which is promoted to a 4D scalar field) is singlet with respect to both SU(2) factors in
 (\ref{globgroup}), i.e. 
the global world-sheet group.\footnote{Which is isomorphic to the 4D
global group \eqref{c+f} at $N=2$, $N_f=4$.} What about its baryonic charge? 

Since
\beq
w_{\alpha}= \frac12\, {\rm Tr}\left[(\bar{\sigma}_{\alpha})_{ KP}\,n^P\rho^K\right]
\label{eq:kinkbaryon}
\eeq
we see that the $b$ state transforms as 
\beq
(1,\,1,\,2),
\label{brep}
\eeq
where we used  (\ref{repsnrho}) and (\ref{deformedconi}). 
Three numbers above refer to the representations of (\ref{globgroup}).
In particular it has the baryon charge $Q_B(b)=2$.

To conclude this section let us note that in our case of type IIA superstring the complex scalar 
associated with deformations of the complex structure of the Calabi-Yau
space enters as a component of a massless 4D \ntwo hypermultiplet, see \cite{Louis} for a review. 
Instead, for type IIB superstring it would be a component of a vector BPS multiplet. Non-vanishing baryonic charge 
of the $b$ state confirms our conclusion that the string under consideration is a type IIA.

\section {Massive states from non-critical \boldmath{$c=1$} string }
\label{c=1}
\setcounter{equation}{0} 

As was explained in Sec. \ref{intro}, the critical string theory on the conifold is hard
to use for calculating the spectrum of massive string modes because the supergravity approximation
does not work. In this section we review the results obtained in \cite{SYlittles} based on the little string theory 
(LST) approach. Namely, in \cite{SYlittles} we used the equivalent formulation of our 
 theory as a non-critical  $c=1$ string theory with the Liouville field and a compact scalar at 
the self-dual radius \cite{GivKut,GVafa}. We intend to use the same formulation in this paper to analyze the
4D hypermultiplet structure of the massive states.

\subsection{Non-critical \boldmath{$c=1$} string theory}

Non-critical $c=1$ string theory is formulated on the target space
\beq
\mathbb{R}^4\times \mathbb{R}_{\phi}\times S^1,
\label{target}
\eeq
where $\mathbb{R}_{\phi}$ is a real line associated with the Liouville field $\phi$ and the theory  
has a linear in $\phi$ dilaton, such that string coupling is given by
\beq
g_s =e^{-\frac{Q}{2}\phi}\, .
\label{strcoupling}
\eeq
We will determine $Q$ in Eq. (\ref{Q}).

Generically the above  equivalence is formulated in the so called double scaling limit between the critical string on non-compact Calabi-Yau spaces with 
an isolated singularity on the one hand, and non-critical $c=1$ string with the additional Ginzburg-Landau
\ntwo superconformal system \cite{GivKut}, on the other hand. Following \cite{GivKut} we assume the double scaling limit when the string coupling constant of the conifold theory $g_{con}$
and the deformation parameter of the conifold $b$ simultaneously go to zero with the combination 
$b^{\frac{Q^2}{2}}/g_{con}$ fixed. In this limit non-trivial physics is localized near the singularity of the Calabi-Yau manifold. In the conifold case the extra Ginzburg-Landau factor 
in \eqref{target} is absent \cite{GivKutP}.

In \cite{ABKS,GivKut,GivKutP} it was argued that non-critical string theories with the string
coupling exponentially falling off at $\phi\to\infty$ are holographic. The string coupling
goes to zero in the bulk of the  space-time  and non-trivial dynamics (LST)\,\footnote{The main 
example of this behavior is non-gravitational LST
in the flat six-dimensional space formed by the world volume of parallel NS5 branes.}  is 
localized on the ``boundary.'' 
In our case the ``boundary'' is the four-dimensional space in which \ntwo SQCD is defined. (It is worth emphasizing that
in our case the boundary 4D dynamics is the starting point while the extra six dimensions represent an auxiliary 
mathematical construct. Perhaps, it can be referred to as a ``reverse holography.")

In other words,  holography for our non-Abelian vortex string theory is most welcome and expected. We start with 
\ntwo SQCD in 4D space and study solitonic vortex strings. In our approach 10D space formed by 4D ``actual''
space and six internal moduli of the string is an artificial construction needed to formulate the string
theory of a special non-Abelian vortex. Clearly we expect that all non-trivial ``actual" physics should 
be localized exclusively on the 
4D ``boundary.'' In other words,  we expect that LST in our case is nothing but 4D
\ntwo SQCD at the critical value of the gauge coupling $g^2_c$ (in the hadronic description).

The linear dilaton in \eqref{strcoupling} implies that the bosonic stress tensor of $c=1$ matter coupled to 
2D gravity is 
\beq
T_{--}= -\frac12\,\left[(\pt_{-} \phi)^2 + Q\, \pt_{-}^2 \phi + (\pt_{-} Y)^2\right], 
\label{T--}
\eeq
where $\pt_{-}=\pt_z$
The  compact scalar $Y$  represents $c=1$ matter and satisfies the following condition:
\beq
Y \sim Y+2\pi Q\, .
\eeq
Here we normalize the scalar fields in such a way that their propagators are
\beq
\langle \phi(z),\phi(0)\rangle = -\log{z\bar{z}}, \qquad \langle Y(z),Y(0)\rangle = -\log{z\bar{z}}\,.
\label{propagators}
\eeq
The central charge of the supersymmetrized $c=1$ theory above is
\beq
c_{\phi+Y}^{SUSY} = 3 + 3Q^2.
\label{cphiY}
\eeq
The criticality condition for the string on \eqref{target} implies  that this central charge should be 
equal to 9. This gives 
\beq
Q=\sqrt{2}\,,
\label{Q}
\eeq
to be used in Eq. (\ref{strcoupling}).

Deformation of the conifold \eqref{deformedconi} translates into adding the Liouville interaction
to the world-sheet sigma model \cite{GivKut},
\beq
\delta L= b\int d^2\theta \, e^{-\frac{\phi + iY}{Q}}\,.
\label{liouville}
\eeq
The conifold singularity at  $b=0$ corresponds to the string coupling constant becoming infinitely large at
$\phi \to -\infty$, see \eqref{strcoupling}. At $b\neq 0$ the Liouville interaction regularize the behavior
of the string coupling preventing the string from propagating to the region of large negative $\phi$.

In fact the $c=1$ non-critical string theory  can also be described in terms of two-dimensional
black hole \cite{Wbh}, which is the ${\rm SL}(2,R)/{\rm U}(1)$ coset WZNW theory 
\cite{MukVafa,GVafa,OoguriVafa95,GivKut} 
at level
\beq
k =\frac{2}{Q^2}\,.
\eeq
In \cite{HoriKapustin} it was shown that \ntwot ${\rm SL}(2,R)/{\rm U}(1)$ coset is a mirror description of the
$c=1$ Liouville theory.
The relation above implies in the case of the conifold ($Q=\sqrt{2}$)  that 
\beq
k=1,
\label{k=1}
\eeq 
where $k$ is the total level of the Ka\v{c}-Moody algebra in the supersymmetric version (the level
of the bosonic part of the algebra is then $k_b=k+2 =3$). The target 
space of 
this theory has the form of a semi-infinite cigar;  the field $\phi$ associated with the motion along the 
cigar
cannot take large negative values due to semi-infinite geometry. In this description the string
coupling constant at the tip of the cigar is $g_s \sim 1/b$. 

In fact as was argued in \cite{GivKut} in the non-critical string theory by itself the  parameter $b$
does not have to be small. If we following \cite{GivKut} take $b$ large the string coupling at the tip of the cigar will be small and the string perturbation theory becomes reliable, cf. \cite{GivKut,Dorey1}. In particular,  we can use the  
tree-level approximaion to obtain the string spectrum.  Note also that as we already mentioned in the Introduction the $SL(2,R)/U(1)$ WZNW model is exactly solvable. 

In terms of 4D SQCD taking $b$ large means moving along the Higgs branch far away from the origin.

\subsection {Vertex operators }
\label{vertexoperators}

Vertex operators for  the  string theory on \eqref{target} are constructed in \cite{GivKut}, see also 
\cite{MukVafa,GivKutP}. Primaries of the $c=1$  part for large 
positive $\phi$ (where the target space becomes a cylinder $\mathbb{R}_{\phi}\times S^1$) take the form
\beq
V^L_{j,m_L}\times V^R_{j,m_R}\approx \exp{\left(\sqrt{2}j\phi + i\sqrt{2}(m_L Y_L +m_R Y_R)\right)},
\label{vertex}
\eeq
where we split $\phi$ and $Y$ into left and right-moving parts, say $\phi=\phi_L +\phi_R$.
 For the self-dual radius \eqref{Q} (or $k=1$) the parameter $2m$ in Eq. (\ref{vertex}) 
is integer. For the
left-moving sector $2m_{L}\equiv 2m$ is 
the total momentum plus the winding number along the compact dimension $Y$. For the right-moving sector
we introduce $2m_{R}$ which is the winding number minus momentum. We will see below 
that for our case type IIA string $m_{R}= - m$, while for type IIB string $m_{R}= m$.

The primary operator \eqref{vertex} is related  to the wave 
function over ``extra dimensions'' as follows:
$$V_{j,m} = g_s \Psi_{j,m}(\phi,Y)\,.$$ The string coupling 
\eqref{strcoupling}  depends on $\phi$. Thus, 
\beq
\Psi_{j,m}(\phi,Y) \sim e^{\sqrt{2}(j+\frac{1}{2})\phi + i\sqrt{2}mY}\,.
\eeq
We will look for string states with normalizable wave functions over the ``extra dimensions'' which we will 
interpret as hadrons in 4D \ntwo SQCD. The 
condition for the string states to have  normalizable wave functions reduces to \footnote{We include 
the case $j=-\frac12$
which is at the borderline between normalizable and non-normalizable states. In \cite{SYlittles} it is shown 
that  $j=-\frac12$  corresponds to the norm logarithmically divergent in the infrared in much the same way 
as the norm of  the $b$ state, see \eqref{Sb}}
\beq
j\le -\frac12\,.
\label{normalizable}
\eeq
The scaling dimension of the primary operator \eqref{vertex} is 
\beq
\Delta_{j,m} = m^2 - j(j+1) \, .
\label{dimV}
\eeq
Unitarity implies that it should be positive,
\beq
\Delta_{j,m}> 0\,.
\label{Deltapositive}
\eeq
Moreover, to ensure that conformal dimensions of left and right-moving parts of the vertex operator
\eqref{vertex} are the same we impose that $m_R=\pm m_L$.

The spectrum of the allowed values of $j$ and $m$ in \eqref{vertex} was  exactly determined by using the Ka\v{c}-Moody algebra
for the coset ${\rm SL}(2,R)/{\rm U}(1)$ in \cite{DixonPeskinLy,Petrop,Hwang,EGPerry,MukVafa}, 
see \cite{EGPerry-rev} for a review. Both discrete and continuous representations were found. Parameters $j$
and $m$ determine the global quadratic Casimir operator and the projection of the spin  on the third axis,
\beq
J^2\, |j,m\rangle\, = -j(j+1)\,|j,m\rangle, \qquad J^3\,|j,m\rangle\, =m \,|j,m\rangle
\eeq
where $J^a$  $(a=1,2,3)$ are the global ${\rm SL}(2,R)$ currents. 

We will focus on  {\em discrete representations} with
\beq
j=-\frac12, -1, -\frac32,..., \qquad m=\pm\{j, j-1,j-2,...\}.
\label{descrete}
\eeq

Discrete representations include the normalizable states localized near the tip of the cigar (see 
\eqref{normalizable}),
 while the continuous representations contain
non-normalizable states. 

Discrete representations contain
states with negative norm. To exclude these ghost states a restriction for spin $j$ is imposed 
\cite{DixonPeskinLy,Petrop,Hwang,EGPerry,EGPerry-rev} 
\beq
 -\frac{k+2}{2}< j <0\,.
\eeq
Thus, for  our value $k=1$ we are left with only two allowed values of $j$,
\beq
j=-\frac12, \qquad m= \pm\left\{\,\frac12,\, \frac32,...\right\}
\label{j=-1/2}
\eeq
and 
\beq
j=-1, \qquad m= \pm\{\,1, \,2,...\}.
\label{j=-1}
\eeq

 Note  that there are also continues (principal and exceptional) representations of primaries of the $c=1$ string theory \cite{EGPerry-rev}, see also a brief  review of discrete and 
continues spectra  in \cite{SYlittles}. In particular, continues representations correspond to  non-normalizable states  in the Liouville direction. Moreover, in \cite{SYlittles} we suggested an interpretation of  these
non-normalizable states: they corresponds to decaying modes of normalizable 4D states. We also confirm this
interpretation showing that spectra of continues states start from thresholds given by masses \eqref{tachyonmass}
and \eqref{gravitonmass} 
of 4D states (see below). Still we believe that the relation between discrete and continues states 
needs future clarification.

\vspace{2mm}

\subsection{Scalar and spin-2 states} 

Four-dimensional spin-0 and spin-2 states were found in \cite{SYlittles} using vertex operators (\eqref{vertex}).
The 4D scalar vertices $V^S$   in the $(-1,-1)$ picture  have the
form \cite{GivKut}
\beq
V^{S,L}_{j,m}\times V^{S,R}_{j,-m}(p_{\mu})= e^{-\varphi_L -\varphi_R }\, e^{ip_{\mu}x^{\mu}}\,
 V^L_{j,m}\times V^R_{j,-m}\, ,
\label{tachyon}
\eeq
where superscript $S$ stands for scalar, $\varphi_{L,R}$ represents bosonized ghost in the left and right-moving sectors, while  $p_{\mu}$ is the 4D momentum of the string state.

The condition for the  state \eqref{tachyon} to be physical is
\beq
\frac12 +\frac{p_{\mu}p^{\mu}}{8\pi T} + m^2 - j(j+1) = 1,
\label{tachphys}
\eeq
where 1/2 comes from the ghost and we used \eqref{dimV}. We note that the conformal dimension
of the ghost operator $\exp{(q\varphi})$ is equal to $-(q+q^2/2)$, where $q$ is the picture number.

The GSO projection restricts the integer $2m$ for the operator in \eqref{tachyon} to be odd
\cite{KutSeib,GivKut} \footnote{We will demonstrate this in the next section.},
\beq
 m=\frac12 +\mathbb{Z}.
\label{oddn}
\eeq
For half-integer $m$ we have only one possibility $j=-\frac12$, see \eqref{j=-1/2}.
This determines the masses of the 4D scalars, 
\beq
\frac{(M^S_{m})^2}{8\pi T}=-\frac{p_{\mu}p^{\mu}}{8\pi T} = m^2 -\frac14 \,,
\label{tachyonmass}
\eeq
where the Minkowski 4D metric with the diagonal entries $(-1,1,1,1)$ is used.

In particular, the state with $m=\pm 1/2$ is the massless baryon $b$, associated with  deformations of the conifold complex structure \cite{SYlittles}, while states with $m = \pm (3/2, 5/2 ,...)$ are massive 4D scalars.

At the next level we consider  4D spin-2 states. 
The corresponding vertex operators are given by
\beq
\left(V^{L}_{j,m}\times V^{R}_{j,-m}(p_{\mu})\right)^{{\rm spin}-2}= 
\xi_{\mu\nu}\psi^{\mu}_L\psi^{\nu}_R\,e^{-\varphi_L -\varphi_R }\, 
e^{ip_{\mu}x^{\mu}}\, V^L_{j,m}\times V^R_{j,-m}\, ,
\label{graviton}
\eeq
where $\psi^{\mu}_{L,R}$ are the world-sheet superpartners to 4D coordinates $x^{\mu}$, while
$\xi_{\mu\nu}$ is the polarization tensor.

The condition for these  states  to be physical takes the form
\beq
\frac{p_{\mu}p^{\mu}}{8\pi T} + m^2 - j(j+1) = 0\,.
\label{gravphys}
\eeq 

The GSO projection selects now $2m$ to be even, $|m|=0, 1,2,...$ \cite{GivKut},
thus we are left with only one allowed value of $j$, $j=-1$ in \eqref{j=-1}. Moreover, the value
$m=0$ is excluded.
This leads to the following expression  for the masses of spin-2 states:
\beq
(M^{{\rm spin}-2}_{m})^2 = 8\pi T\,m^2, \qquad |m|=1,2,... .
\label{gravitonmass}
\eeq
We see that all spin-2 states are massive. This confirms the result in \cite{KSYconifold} that 
no massless 4D graviton appears in our theory. It also matches the fact that our ``boundary'' theory, 4D 
\ntwo QCD, is defined in flat space without gravity.

\begin{figure}
\epsfxsize=10cm
\centerline{\epsfbox{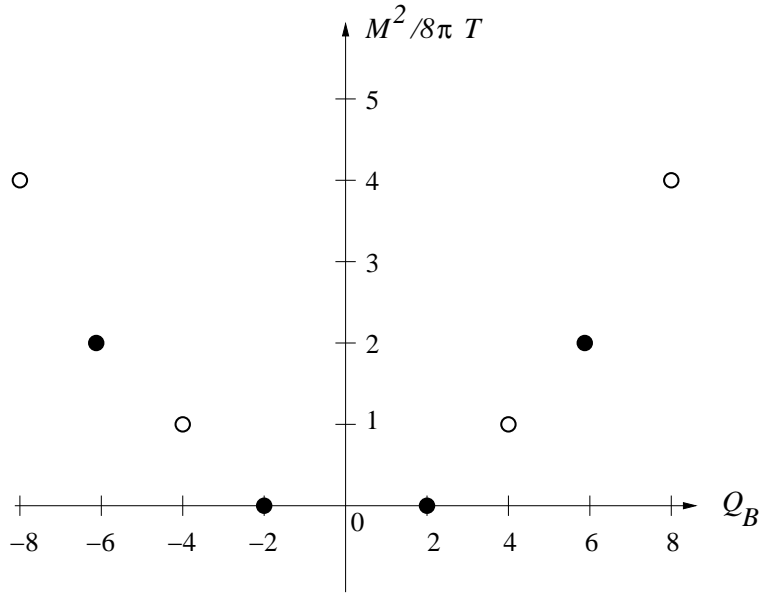}}
\caption{\small  Spectrum of spin-0 and spin-2 states as a function of the baryonic charge. Closed 
and open circles denote  spin-0 and spin-2 states, respectively.
 }
\label{fig_spectrum}
\end{figure}

To determine baryonic charge of these states we note that U(1)$_B$ transformation of $b$ in 
the Liouville interaction \eqref{liouville} is compensated by a shift of $Y$. The baryonic charge of $b$ is  two, see \eqref{brep}. Below we use the following convention: upon splitting $Y$ into left and right-moving
parts $Y=Y_L+Y_R$ we define that only $Y_L$ is shifted under U(1)$_B$ transformation,
\beq
b\to e^{2i\theta}b, \qquad Y_L\to Y_L +2\sqrt{2}\theta, \qquad Y_R\to Y_R.
\label{baryonshift}
\eeq
This gives for the baryon charge of the vertex operator \eqref{vertex} 
\beq
 Q_B = 4m.
\label{m-baryon}
\eeq
We see that the momentum $m$ in the compact $Y$ direction is in fact the baryon charge of a string state. All states we found above are baryons. Their  masses  as a function of the 
baryon charge are shown in Fig.~\ref{fig_spectrum}.

The momentum $m$ in the compact dimension is also related to the $R$-charge. On the world sheet we can 
introduce the left and right $R$-charges separately. Normalizing  charge of $\theta^{+}$, namely, 
$R^{(2)}_L (\theta^{+}) =1$, we see that  
$Y$ should be shifted under the $R^{(2)}_L$ symmetry to make  invariant the Liouville interaction \eqref{liouville}.

This gives 
\beq
R^{(2)}_L(V^L_{j,m}) = -2m
\label{R_L}
\eeq
for the $R^{(2)}_L$ charge of the vertex \eqref{vertex} which is the bottom component of the world sheet
supermultiplet. The $R^{(2)}_R$ charge in the right-moving sector is defined similarly. Here superscript $(2)$ denotes  the world sheet $R$-charge. 

As was discussed above, the massless baryon $b$ corresponds to $j=-1/2$, $m=\pm 1/2$. Thus, the  associated vertex 
$V_{j,m}$ has $R^{(2)}_L= \pm 1$ and conformal dimension $\Delta = 1/2$,
see \eqref{dimV}. Therefore it satisfies the relation 
\beq
\Delta =\frac{|R^{(2)}_L|}{2}
\label{chiral} 
\eeq
as expected for the bottom component of a chiral primary operator, which defines the  short representation of supersymmetry algebra (and similar relation in the right-moving sector). In 4D theory $b$ is a component of a short \ntwo BPS multiplet, namely hypermultiplet.

\section {Massless hypermultiplet }
\label{b}
\setcounter{equation}{0} 

The remainder of this paper is devoted to the  study the supermultiplet structure of the 4D string states described  in the previous sections. Our strategy is as follows: we explicitly construct 4D supercharges and use them to generate all components  of a given multiplet starting from a scalar or spin-2 representative  shown in \eqref{tachyon} or 
\eqref{graviton}. We will generate supermultiplets originating from the lowest states with $j=-1/2$, $m=\pm(1/2,3/2)$ and
$j=-1$, $m=\pm 1 $. In this section we will start with the massless baryon $b$.

\subsection{4D supercharges}

First we bosonize world sheet fermions $\psi_{\mu}$, $\psi_{\phi}$ and 
$\psi_{Y}$, the superpartners of $x_{\mu}$, the Liouville field $\phi$ and the compact scalar $Y$, respectively.
Following the standard rule we divide them into pairs
\beq
\psi_k = \frac{1}{\sqrt{2}}(\psi_{2k-1}-i\psi_{2k}), \qquad 
\bar{\psi}_k = \frac{1}{\sqrt{2}}(\bar{\psi}_{2k-1}+i\bar{\psi}_{2k}), \qquad k=1,2,
\label{psik}
\eeq
\beq
\psi = \frac{1}{\sqrt{2}}(\psi_{\phi}-i\psi_{Y}), \qquad 
\bar{\psi} = \frac{1}{\sqrt{2}}(\bar{\psi}_{\phi}+i\bar{\psi}_{Y}), 
\label{psi}
\eeq 
and define 
\beq
\psi_k \bar{\psi}_k =i\pt_{-}H_k \qquad {\rm (no\,\, summation)}, \qquad \psi\bar{\psi} =i\pt_{-}H,
\label{H}
\eeq
where the bosons $H_k$ and $H$ have the standard propagators 
\beq
\langle H_k(z),H_l(0)\rangle = -\delta_{kl}\log{z}, \qquad \langle H(z),H(0)\rangle = -\log{z}
\eeq
and 
\beq
\psi_k \sim e^{iH_k}, \qquad \psi \sim e^{iH}.
\label{psiH}
\eeq
The above formulas are written for the left-moving sector. In the right-moving sector bosonization is similar with the replacement  $ z \to \bar{z}$ and $\pt_{z} \to \pt_{\bar{z}}$.

As usual, we define spinors in terms of scalars $H$.  Namely,
\beq
S_{\alpha} = e^{\sum_k is_k H_k}, \qquad \bar{S}_{\dot{\alpha}} = e^{\sum_k i\bar{s}_k H_k}
\label{4Dspinors}
\eeq
are  4D spinors, $\alpha=1,2$, ${\dot{\alpha}} = 1,2$.
Moreover,
\beq
S = e^{ i\frac{ H}{2}}, \qquad \bar{S}= e^{ -i\frac{ H}{2}}
\label{spinor}
\eeq
are spinors associated with '``extra'' dimensions $\phi$ and $Y$. Here $s_k =\pm \frac12$, $k=1,2$ and
the choices of the allowed values of $s_k$ are restricted by the GSO projection, see below.

Supercharges for non-critical string are defined in \cite{KutSeib}. In  our case four 4D \none supercharges
\beqn
Q_{\alpha} &=& \frac{1}{2\pi i}\, \frac{\bar{b}}{|b|} \int d z \,e^{-\frac{\varphi}{2}} S_{\alpha}\, S\, 
\exp{\left(\frac{i}{\sqrt{2}}Y\right) },
\nonumber\\[3mm]
\bar{Q}_{\dot{\alpha}} &=& \frac{1}{2\pi i}\, \frac{b}{|b|} \int d z \,e^{-\frac{\varphi}{2}} \bar{S}_{\dot{\alpha}}\, \bar{S} \,
\exp{\left(-\frac{i}{\sqrt{2}}Y \right)}
\label{supercharges}
\eeqn
act in the left-moving sector, where we used the $\left(-\frac 12\,\right)$ picture. We have to multiply these supercharges
in the left-moving sector by the
phase factors $\bar{b}/|b|$ and $b/|b|$  to make them neutral with respect to baryonic U(1)$_B$. Other four supercharges of \ntwo 4D supersymmetry are given by similar formulas and act in the right-moving sector. The action of the supercharge on a vertex
is understood as an integral around the location of the vertex on the world sheet.

Supercharges \eqref{supercharges} satisfy 4D space-time supersymmetry algebra 
\beq
\{Q_{\alpha},\bar{Q}_{\dot{\alpha}}\} = 2P_{\mu}\sigma^{\mu},
\label{algebra}
\eeq
while all other anti-commutators vanish. Note that $P_{\mu}$ is the 4D momentum operator, the anti-commutator
\eqref{algebra} does not produce translation in the Liouville direction.

The GSO projection is the requirement of locality of a given vertex operator with respect to the  supercharges
\eqref{supercharges}. 

Let us start with $Q_{\alpha}$ with $s^{(0)}_k= (1/2,\,1/2)$. Then mutual  locality
of the supercharges \eqref{supercharges} selects polarizations
\beq
s_k= \pm \left(\frac12,\, \frac12\right), \qquad \bar{s}_k= \pm \left(\frac12,\,-\frac12\right)
\label{s^0}
\eeq
associated with four supercharges $Q_{\alpha}$ and $\bar{Q}_{\dot{\alpha}}$.

As an  example, let us check the GSO selection rule \eqref{oddn} for 10D '``tachyon''
vertices \eqref{tachyon}.
We have
\beq
\langle Q_{\alpha}, V^{S,L}_{jm}(w) \rangle \sim \int  dz \,\left\{(z-w)^{-(\frac12 - m)} +...\right\}
\eeq
where dots stand for less singular OPE terms and $1/2$ comes from the ghost $\varphi$. We see that locality requirement selects half-integer $m$ as shown in 
\eqref{oddn}. Note that an important feature of the supercharges \eqref{supercharges} is 
the dependence on momentum $m$
in the compact direction $Y$. Without this dependence all 10D ``tachyon'' vertices \eqref{tachyon} would be 
projected out as it happens for critical strings. Note also that none of the states \eqref{tachyon} are 
tachyonic in 4D.

Now we can introduce 4D space-time $R$-charges. We normalize them as follows:
\beq
R^{(4)} = R^{(4)}_L + R^{(4)}_R, \qquad R^{(4)}_L(Q_{\alpha}) =-1, \qquad R^{(4)}_L(\bar{Q}_{\dot{\alpha}}) =1\,,
\eeq
and use the same normalizations for $R^{(4)}_R$.
This definition ensures that for a given vertex operator we have
\beq
R^{(4)}_L =-2m_L, \qquad R^{(4)}_R =-2m_R.
\label{R4}
\eeq
Note that the scalars $H$ are not shifted upon $R^{(4)}$ rotations, so the world-sheet fermions $\psi_k$, $\psi$ do not have $R^{(4)}$
charges. This is in contrast with the action of  the world sheet $R^{(2)}$ symmetry. 

\subsection{Fermion vertex}

To generate fermion vertex for the $b$ state we apply supercharges \eqref{supercharges} to the left-moving
part of the vertex \eqref{tachyon} with $j=-1/2$ and  $m=\pm 1/2$. To get the fermion vertex in the standard $(-1/2)$
picture we have to convert the vertex \eqref{tachyon} from the $(-1)$ to (0) picture. This is done in Appendix A using the 
BRST operator. The  left-moving part of the scalar vertex \eqref{tachyon} in the (0) picture has the form
\beq
V^{(0)}_{j,m}(p_{\mu})= \left[\sqrt{2}(j\psi_{\phi}+im\psi_{Y}) + \frac{i}{\sqrt{4\pi T}}\,p_{\mu}\psi^{\mu}\right] \,
e^{ip_{\mu}x^{\mu} +\sqrt{2}j\phi + i\sqrt{2} m Y}\, ,
\label{tachyon0}
\eeq
where we skip the subscripts $L$.

\vspace{1mm}

Let us start with  $j=-1/2$ and  $m = 1/2$. The vertex \eqref{tachyon0} reduces to 
\beq
V^{(0)}_{-\frac{1}{2},\,m=\frac12}(p_{\mu})= \left[-\psi + \frac{i}{\sqrt{4\pi T}}\,p_{\mu}\psi^{\mu}\right] \,
e^{ip_{\mu}x^{\mu} - \frac{\phi}{\sqrt{2}} + i \frac{ Y}{\sqrt{2}}}\, .
\label{tachyonb}
\eeq
 Applying the supercharge $Q_{\alpha}$ we find that correlation function does not contain
pole contribution and hence gives zero. On the other hand $\bar{Q}_{\dot{\alpha}}$ produces the following fermion vertex
\beqn
\bar{V}_{\dot{\alpha}}^{(-\frac12)} &=& \langle \bar{Q}_{\dot{\alpha}}, V^{(0)}_{-\frac12,\, m=\frac12}(p_{\mu})\rangle 
\nonumber\\[3mm]
 &\sim & e^{-\frac{\varphi}{2}}\left[-\bar{S}_{\dot{\alpha}} \,S + \frac{ip_{\mu}}{\sqrt{4\pi T}}\, 
(\bar{\sigma}_{\mu})_{\dot{\alpha}\alpha} S^{\alpha}\, \bar{S}\right] \,e^{ip_{\mu}x^{\mu} 
- \frac{\phi}{\sqrt{2}}}
\label{fermionVdot1/2}
\eeqn
where we used 
\beqn
& & \langle \psi (z), \bar{S}(w)\rangle \sim \frac{1}{\sqrt{(z-w)}}\,S\,,
\nonumber\\[3mm]
& & \langle e^{\frac{iY(z)}{\sqrt{2}}}, e^{\frac{-iY(w)}{\sqrt{2}}}\rangle \sim \frac{1}{\sqrt{(z-w)}}\,,
\nonumber\\[3mm]
& & \langle \psi_{\mu} (z), \bar{S}(w)_{\dot{\alpha}}\rangle \sim \frac{1}{\sqrt{(z-w)}}\,
(\bar{\sigma}_{\mu})_{\dot{\alpha}\alpha} S^{\alpha}\,.
\label{correlators}
\eeqn
Note that the momentum $m$ along the compact direction is zero for the fermion vertex \eqref{fermionVdot1/2}.

As a check we can calculate the conformal dimension of the vertex \eqref{fermionVdot1/2}. The condition for this vertex  
 to be physical is
\beq
\frac38 +\frac38 +\frac{p_{\mu}p^{\mu}}{8\pi T} - j(j+1) = 1\,,
\label{fermdimension}
\eeq
where the first and the second contributions come from the ghost $\varphi$ and the scalars $H_k$ and $H$,
respectively. We see that for $j=-1/2$ this state is massless, as expected.

By the same token, for $m= -1/2$ we consider the action of the supercharges on the vertex in  
\eqref{tachyonb} with $\psi \to \bar{\psi}$ and $m=-1/2$.
Only the action of $Q_{\alpha}$ gives non-trivial fermion vertex. We get
\beqn
V^{\alpha,(-\frac12)} &=& \langle Q^{\alpha}, V^{(0)}_{-\frac12,\,m=-\frac12}(p_{\mu})\rangle 
\nonumber\\[3mm]
 &\sim & e^{-\frac{\varphi}{2}}\left[-S^{\alpha} \,\bar{S} + \frac{ip_{\mu}}{\sqrt{4\pi T}}\, 
(\sigma_{\mu})^{\alpha\dot{\alpha}} \bar{S}_{\dot{\alpha}}\, S\right] \,e^{ip_{\mu}x^{\mu} 
- \frac{\phi}{\sqrt{2}}}\,.
\label{fermionV1/2}
\eeqn

To conclude this subsection we note that if we apply supercharges to the fermion vertices \eqref{fermionVdot1/2}
and \eqref{fermionV1/2} we do not generate new states. For example, acting on \eqref{fermionVdot1/2} with
$Q_{\alpha}$ gives (the left-moving part of) the scalar vertex \eqref{tachyon},
\beq
\langle Q_{\alpha}, \bar{V}_{\dot{\alpha}}^{(-\frac12)}\rangle \sim \frac{p_{\mu}}{\sqrt{4\pi T}}\,
(\bar{\sigma}_{\mu})_{\dot{\alpha}\alpha}\,V^{S,L}_{-\frac12,\,m=\frac12}
\eeq
in the picture $(-1)$. This result is in full accord with supersymmetry algebra \eqref{algebra}. Acting with
$Q_{\dot{\alpha}}$ produces the scalar vertex \eqref{tachyon} with $m=-1/2$,
\beq
\langle Q_{\dot{\alpha}}, \bar{V}_{\dot{\beta}}^{(-\frac12)}\rangle \sim \varepsilon_{\dot{\alpha}\dot{\beta}}
\, V^{S,L}_{-\frac12,\,m=-\frac12}.
\eeq

\subsection{Building the hypermultiplet}

In this section we will use the bosonic and fermionic vertices obtained above to construct hypermultiplet of the massless 
$b$ states. For simplicity in this section and below we will consider only bosonic components of supermultiplets.
As was already mentioned,  in the case  of type IIA superstring
we should consider the states with $m_R=-m_L \equiv -m$. We will prove this statement below, in this and the subsequent subsections.

In the NS-NS sector we have one complex (or two real) scalars \eqref{tachyon},
\beq
b= V^{S,L}_{j=-\frac12,\,m}\times V^{S,R}_{j=-\frac12,\,-m}
\label{bNSNS}
\eeq
associated with $m=\pm 1/2$.

Since for the scalar states the momentum $m$ is opposite in the left- and right- moving sectors, for the R-R states we get the
product of fermion vertices \eqref{fermionVdot1/2} and \eqref{fermionV1/2}, namely,
\beq
V_{\dot{\alpha}\alpha} = \bar{V}^L_{\dot{\alpha}} \times V^R_{\alpha}, \qquad 
\bar{V}_{\alpha\dot{\alpha}} = V^L_{\alpha} \times \bar{V}^R_{\dot{\alpha}}. 
\label{bRR}
\eeq
The vertices above define a complex vector $C^{\mu}$ via
\beq
V_{\dot{\alpha}\alpha} = (\bar{\sigma}_{\mu})_{\dot{\alpha}\alpha} C^{\mu}\,.
\label{bvector}
\eeq
However, as is usual for the massless R-R string states, the number of physical degrees of freedom reduces because 
the fermion vertices \eqref{fermionVdot1/2} and \eqref{fermionV1/2} satisfy the massless Dirac equations which translate 
into the Bianchi identity for the associated form. For 1-form (vector) we have
\beq
\pt_{\mu} C_{\nu} - \pt_{\nu} C_{\mu}=0\,,
\label{bianchi}
\eeq
which ensures that the complex vector reduces to a complex scalar,
\beq
C_{\mu} = \pt_{\mu} \tilde{b}\, .
\label{tildeb}
\eeq

Altogether we have two complex scalars, $b$ and $\tilde{b}$, which form the bosonic part of the hypermultiplet.
As was already mentioned, deformations of the complex structure of a Calabi-Yau manifold gives a massless hypermultiplet for type IIA theory and massless vector multiplet for type IIB theory. The  derivation above shows 
that our choice $m_R=-m_L$ corresponds to type IIA string. 

We stress again that this massless hypermultiplet is a short BPS representation of \ntwo supersymmetry algebra
in 4D and is characterized by the non-zero baryonic charge $Q_B(b)=\pm 2$.

Let us also note that the four-dimensional 
space-time $R^{(4)}$ charge of the vertex operator \eqref{bNSNS} vanishes due to cancellation between left and 
right-moving sectors, see \eqref{R4}. For the vertex \eqref{bRR} it is also zero since both $m_L$ and $m_R$ are zero.
Thus we conclude that $b$ and $\tilde{b}$ have the vanishing  $R^{(4)}$ charge, as expected for the scalar components of a hypermultiplet.

\subsection{What would we get for type IIB superstring?}

Our superstring is of type IIA. This is fixed by derivation of our string theory
as a description of non-Abelian vortex in 4D \ntwo SQCD, see \cite{KSYconifold}. In this subsection we ``forget'' for a short while about this and consider superstring theory on the manifold \eqref{target} on its own right. Then, as usual in string theory,
we have two options for a closed string: type IIA and type IIB. We will show below that type IIB option
corresponds to the choice $m_R=m_L$.

For this choice  the massless state with $j=-1/2$ is described as follows. In the NS-NS sector 
we have  one complex scalar,
\beq
a= V^{S,L}_{j=-\frac12,\,m}\times V^{S,R}_{j=-\frac12,\,m}\,,
\label{aNSNS}
\eeq
associated with $m=\pm 1/2$. In the R-R sector we now obtain
\beq
V_{\alpha\beta} = V^L_{\alpha} \times V^R_{\beta}, \qquad 
\bar{V}_{\dot{\alpha}\dot{\beta}} = \bar{V}^L_{\dot{\alpha}} \times \bar{V}^R_{\dot{\beta}}. 
\label{aRR}
\eeq
Expanding the complex vertex $V_{\alpha\beta}$ in the basis of $\sigma$ matrices 
\beq
V^{\alpha}_{\beta} = F \,\delta _{\alpha}^{\beta} + 
(\sigma_{\mu}\bar{\sigma}_{\nu})^{\alpha}_{\beta} \,C^{\mu\nu}
\label{IIB2form}
\eeq
we get a complex scalar $F$ and a complex  2-form $C_{\mu\nu}$ which can be expressed in terms of a real 2-form, 
$C_{\mu\nu} = F_{\mu\nu} - i F_{\mu\nu}^{*}$,
where $F_{\mu\nu}$ is real and 
$F_{\mu\nu}^{*} = \frac12\,\varepsilon_{\mu\nu\rho\lambda}F^{\rho\lambda}$. The Dirac equations for the fermion vertices 
\eqref{fermionVdot1/2} and \eqref{fermionV1/2}
imply that $F$ is a constant, while $F_{\mu\nu}$ satisfies the Bianchi identity. This ensures that $F_{\mu\nu}$ can
be constructed in terms of a real vector potential
\beq
F_{\mu\nu} = \pt_{\mu}A_{\nu} -\pt_{\nu} A_{\mu}.
\eeq

We see that we get  a massless \ntwo BPS vector multiplet with the bosonic components given by the complex scalar 
$a$ and the gauge potential $A_{\mu}$. This is what we expect from deformation of the complex structure 
of a Calabi-Yau manifold for  type IIB string.

Let us  note that $R$ charges also match since the $R^{(4)}$ charge of $a$ in \eqref{aNSNS} is $R^{(4)}= \pm 2$
(see \eqref{R4}) while the $R^{(4)}$ charge of \eqref{aRR} and $A_{\mu}$ are zero as expected.

However, if we try to interpret this \ntwo vector multiplet as a state of the non-Abelian vortex
in \ntwo SQCD we will get an inconsistency. To see this one can observe that our state has non-zero baryonic charge
which cannot be associated with a gauge multiplet. This confirms our conclusion that the string theory
for our non-Abelian vortex-string is of IIA type.

\section{Exited state with \boldmath{$j=-1/2$}}
\label{m=3/2}
\setcounter{equation}{0} 

Below  we consider the supermultiplet structure of the lowest massive states given by the vertex operators
\eqref{tachyon} and \eqref{graviton}. In this section we start with the first excited 
state of the scalar vertex \eqref{tachyon} with $j=-1/2$ and $m=\pm 3/2$. The mass of this state is 
\beq
\frac{\left(M_{j=-\frac12, m=\pm 3/2}\right)^2}{8\pi T}= 2 \,,
\label{mass3/2}
\eeq
see \eqref{tachyonmass}.

\subsection{Action of supercharges}

The left-moving part of the vertex operator in the (0) picture is 
given by \eqref{tachyon0}. For  $m=3/2$ we obtain
\beq
V^{(0)}_{-\frac{1}{2},\,\frac32}(p_{\mu})= \left[-(2\psi -\bar{\psi})+ \frac{i}{\sqrt{4\pi T}}\,p_{\mu}\psi^{\mu}\right] \,
e^{ip_{\mu}x^{\mu} - \frac{\phi}{\sqrt{2}} + i\frac{3}{\sqrt{2}}\,  Y}\, .
\label{tachyon3/2}
\eeq
In much the same way as for the $b$ state, the supercharge $Q$ acting on the vertex above gives zero while the
supercharge $\bar{Q}$ produces the following fermion vertex in the picture $\left(-\frac 12\right)$:
\beqn
&&\bar{V}_{\dot{\alpha}}^{(-\frac12)} = \langle \bar{Q}_{\dot{\alpha}}, V^{(0)}_{-\frac12,\, m=\frac32}(p_{\mu})\rangle \sim e^{-\frac{\varphi}{2}}\left[-2\bar{S}_{\dot{\alpha}} \,S
\right.
\nonumber\\[3mm]
&&
\left.
  + \frac{ip_{\mu}}{\sqrt{4\pi T}}\, 
(\bar{\sigma}_{\mu})_{\dot{\alpha}\alpha} S^{\alpha}\, \bar{S}\right] (\pt_{-} Y + \psi_{\phi}\psi_{Y})
\,e^{ip_{\mu}x^{\mu}  - \frac{\phi}{\sqrt{2}}+ i\sqrt{2} Y}.
\label{fermionVdot3/2}
\eeqn
Note that the momentum $m$ along the compact dimension is $$m=1$$ for this vertex. It is easy to check that the mass of this fermion is given by \eqref{mass3/2}.

In a similar manner, for $m=-3/2$ we use the bosonic vertex \eqref{tachyon3/2} with $\psi \to \bar{\psi}$ and $m=-3/2$.
Action of supercharge $Q$ gives the following fermion vertex: 
\beqn
&& V^{\alpha,(-\frac12)} = \langle Q^{\alpha}, V^{(0)}_{-\frac12,\,m=-\frac12}(p_{\mu})\rangle 
\sim  e^{-\frac{\varphi}{2}}\left[-2S^{\alpha} \,\bar{S}
\right.
\nonumber\\[3mm]
 && 
\left.
+ \frac{ip_{\mu}}{\sqrt{4\pi T}}\, 
(\sigma_{\mu})^{\alpha\dot{\alpha}} \bar{S}_{\dot{\alpha}}\, S\right] (\pt_{-} Y + \psi_{\phi}\psi_{Y})
\,e^{ip_{\mu}x^{\mu} - \frac{\phi}{\sqrt{2}} - i\sqrt{2} Y}
\label{fermionV3/2}\,,
\eeqn
with $m=-1$.

Now let us apply the supercharges to the  fermion vertices \eqref{fermionVdot3/2} and \eqref{fermionV3/2}. Action
of $Q$ on \eqref{fermionVdot3/2} does not produce new states, while $\bar{Q}$ gives 
\beq
\langle \bar{Q}_{\dot{\alpha}}, \bar{V}_{\dot{\beta}}^{(-\frac12)} \rangle \sim 
\varepsilon_{\dot{\alpha}\dot{\beta}}\, V^{S,\,{\rm excited}}_{m=\frac12},
\eeq
where the new excited scalar vertex in the picture $(-1)$ with $m=1/2$ has the form
\beq
V^{S,\,{\rm excited}}_{m=\frac12}= \left[ -2 \pt_{-}^2 Y
 + \frac{ip_{\mu}}{\sqrt{\pi T}}\,\psi^{\mu}\bar{\psi}\, \pt_{-}Y\right] \,e^{-\varphi}
\,e^{ip_{\mu}x^{\mu}  - \frac{\phi}{\sqrt{2}}+ i \frac{Y}{\sqrt{2}}}.
\label{newscalar3/2}
\eeq
The mass of this state is still given by \eqref{mass3/2}. Action of supercharge $Q$ on the fermion
vertex \eqref{fermionV3/2} produces the conjugated scalar with $m=-1/2$.

\subsection{Building massive vector supermultiplet}

Now we can use the vertices obtained in the previous subsection to construct supermultiplets at the level
\eqref{mass3/2}. We have two scalar vertices with $m=\pm3/2$ and $m=\pm1/2$, see left-hand side of 
\eqref{tachyon} and  \eqref{newscalar3/2}. Using these vertices we can construct the scalar states in the NS-NS sector.
Namely, we have one complex scalar 
\beq
V^{S,L}_{j=-\frac12,\,m=\pm \frac32}\times V^{S,R}_{j=-\frac12,\,m=\mp \frac32}
\label{NSNS3/2}
\eeq
formed by the $m=\pm 3/2$ vertices and one complex scalar 
\beq
V^{S,\,{\rm excited},L}_{j=-\frac12,\,m=\pm \frac12}\times V^{S,\,{\rm excited},R}_{j=-\frac12,\,m=\mp \frac12}
\label{newNSNS3/2}
\eeq
 formed by the $m=\pm 1/2$ vertices \eqref{newscalar3/2}.

Moreover, we have also another two complex scalars,
\beq
V^{S,L}_{j=-\frac12,\,m=\pm \frac32}\times V^{S,\,{\rm excited},R}_{j=-\frac12,\,m=\mp \frac12}
\label{cross3/2}
\eeq
and 
\beq
V^{S,\,{\rm excited},L}_{j=-\frac12,\,m=\pm \frac12}\times V^{S,R}_{j=-\frac12,\,m=\mp \frac32}
\label{cross3/2bar}
\eeq
formed by products of two different vertices. Altogether in the NS-NS sector we observe four complex scalars.

In the R-R sector we have 
\beq
V_{\dot{\alpha}\alpha}^{{\rm excited}} = \bar{V}^L_{\dot{\alpha}} \times V^R_{\alpha}, \qquad 
\bar{V}_{\alpha\dot{\alpha}}^{{\rm excited}} = V^L_{\alpha} \times \bar{V}^R_{\dot{\alpha}}, 
\label{RR3/2}
\eeq
where now the fermion vertices are given by \eqref{fermionVdot3/2} and \eqref{fermionV3/2}.
Expanding these vertices in the basis of $\sigma$ matrices 
\beq
V_{\dot{\alpha}\alpha}^{{\rm excited}} = (\bar{\sigma}_{\mu})_{\dot{\alpha}\alpha} B^{\mu}
+(\bar{\sigma}_{\mu}\sigma_{\nu}\bar{\sigma}_{\rho})_{\dot{\alpha}\alpha} B^{\mu\nu\rho}
\label{expansion3/2}
\eeq
we arrive at the complex vector field $B^{\mu}$ and the complex 3-form $B^{\mu\nu\rho}$.

In four dimensions the massive 3-form is dual  to a massive scalar \cite{massivePform}.\footnote{We did not include 3-form in 
the expansion  \eqref{bvector} because in the massless case it contains no physical degrees of freedom, see below.}
Generically the rules of dualizing can be summarized as follows \cite{massivePform}. In $D$ dimensions
massless $p$-forms have 
\beq
c^p_{D-2} = \frac{(D-2)!}{p!(D-2-p)!}
\eeq
physical degrees of freedom. Therefore, the rule of dualizing of the massless $p$-form is
\beq
p \rightarrow (D-2-p).
\label{dualm=0}
\eeq
In particular, 3-form in 4D has no degrees of freedom.

For the massive $p$ forms we have 
\beq
c^p_{D-1} = \frac{(D-1)!}{p!(D-1-p)!}
\eeq
physical degrees of freedom. The rule of dualizing now becomes 
\beq
p \rightarrow (D-1-p).
\label{dualm>0}
\eeq

Thus the massive 3-form in 4D is dual to a massive scalar. Explicitly the duality relation can be written as 
\cite{massivePform,Bukhbinder}
\beq
 B_{\mu\nu\rho}\sim \varepsilon_{\mu\nu\rho\lambda}\, \pt^{\lambda} c\,.
\eeq

We conclude  in the R-R sector we obtained one complex scalar $c$ and the complex vector $B^{\mu}$.
Altogether the bosonic part of the supermultiplet with mass \eqref{mass3/2} contains 5 scalars and 
a vector, all complex. This is exactly the bosonic content of two real \ntwo long massive vector multiplets, each  containing
5 scalars and a vector, see Appendix B,
\beq
({\mathcal N}=2\;)_{ {\rm vector}} =   1_{{\rm vector}} + 5_{{\rm scalar}}\,.
\label{N=2vector}
\eeq 

Let us note that the \ntwo  massive vector multiplet can be realized 
as a result of Higgsing of a $U(1)$ massless gauge multiplet containing gauge field and a complex scalar (2 real
scalars) by vacuum expectation values (VEVs) of a hypermultiplet which contains 4 real scalars. After Higgsing, one scalar is ``eaten'' by the Higgs mechanism, so we are left with massive vector field and 5 scalars. The number of
degrees of freedom in this massive \ntwo long vector multiplet is 8=3+5, where 3 comes from the massive vector.

Summarizing this section we present 4D $R$ charges of the vector multiplet components. Due to cancellation of 
the $R$ charges of the left and right-moving sectors, the $R^{(4)}$ charges of the R-R states \eqref{RR3/2}
and two scalars \eqref{NSNS3/2}, \eqref{newNSNS3/2} of the NS-NS sector vanish, see \eqref{R4}. The $R$-charges
of two scalars \eqref{cross3/2} and \eqref{cross3/2bar} are non-zero, $R^{(4)}=\pm 2$. These are exactly the $R$-charges of a massive \ntwo vector multiplet.
This can be easily understood in terms of Higgsing of the massless gauge multiplet by hypermultiplet VEVs.
The gauge field and scalars from the hypermultiplet have the zero $R$ charge while the $R$ charges of two
scalar superpartners of the gauge field in the massless vector multiplet are indeed characterized by $R^{(4)}=\pm 2$,
cf. Sec. 4.4.

\section {The lowest \boldmath{$j=-1$} multiplet }
\label{j=1}
\setcounter{equation}{0} 

In this section we consider the lowest spin-2 supermultiplet produced by the vertex operator \eqref{graviton}. 
 The mass of the state with $j=-1$ and $m=\pm 1$ is 
\beq
\frac{\left(M_{j=-1, m=\pm 1}\right)^2}{8\pi T}= 1 \,,
\label{mass1}
\eeq
see \eqref{gravitonmass}.

We will
see below that the spin-2 state \eqref{graviton} is the highest component of this supermultiplet. To simplify
our discussion it is easier to start from a scalar component of this supermultiplet replacing the world-sheet
fermions $\psi_{\mu}^{L,R}$ by $\psi_{\phi}^{L,R}$ and $\psi_Y^{L,R}$. Thus, in the left-moving sector
we start from the scalar vertex which, in the picture $(-1)$,  has the form
\beq
V_{j=-1,m= 1}^{(-1)} = \psi\,e^{-\varphi}\,
e^{ip_{\mu}x^{\mu} - \sqrt{2}\phi  + i\sqrt{2}m  Y}\,
\label{scalarj=-1}
\eeq
where  we skip the superscripts $L$, while $\psi$ is given by \eqref{psi} and $m= 1$. 
For $m=-1$ we use a similar vertex with replacement $\psi \to \bar{\psi}$. The conformal dimension of this vertex is the same
as that of the vertex in \eqref{graviton}, so we have a scalar state with mass \eqref{mass1}.

\subsection{Action of supercharges}

To convert this vertex operator into the picture $(0)$ we use the BRST operator, see Appendix A. Then in the picture (0) we
have
\beq
V_{j=-1,m= 1}^{(0)} = \left[ \frac{1}{\sqrt{2}}(\pt_{-}\phi -i\pt_{-} Y)
+ \frac{ip_{\mu}}{\sqrt{4\pi T}}\,\psi^{\mu}\psi\right]\,e^{-\varphi}\,
e^{ip_{\mu}x^{\mu} - \sqrt{2}\phi  + i\sqrt{2}m  Y}\,
\label{picture0j=-1}
\eeq
for $m=1$ and a similar vertex with $\psi \to \bar{\psi}$ for $m=-1$.

Now, let us apply the supercharges to generate the fermion vertices. $Q$ acts trivially on \eqref{picture0j=-1}, 
while $\bar{Q}$
produces the following fermion vertex in the  picture $(-1/2)$:
\beqn
&&\bar{V}_{\dot{\alpha}}^{(-\frac12)} = \langle \bar{Q}_{\dot{\alpha}}, V^{(0)}_{-1,\, m=1}(p_{\mu})\rangle 
\sim e^{-\frac{\varphi}{2}}\left[\bar{S}_{\dot{\alpha}} \,\bar{S}
\right.
\nonumber\\[3mm]
&&
\left.
  + \frac{p_{\mu}}{\sqrt{4\pi T}}\, 
(\bar{\sigma}_{\mu})_{\dot{\alpha}\alpha} S^{\alpha}\, S\right] (\pt_{-} Y + \psi_{\phi}\psi_Y)
\,e^{ip_{\mu}x^{\mu}  - \sqrt{2}\phi+ i \frac{Y}{\sqrt{2}}},
\label{fermionVdot1}
\eeqn
 This fermion vertex has $m=1/2$. 

In a similar manner applying supercharge $Q$ to the scalar vertex $V_{j=-1,m= -1}^{(0)}$ we get
 a fermion vertex with $m=-1/2$,
\beqn
&& (V^{\alpha})^{(-\frac12)} = \langle Q^{\alpha}, V^{(0)}_{-1,\, m= -1}(p_{\mu})\rangle 
\sim e^{-\frac{\varphi}{2}}\left[ S^{\alpha}\, S
\right.
\nonumber\\[3mm]
&&
\left.
  + \frac{p_{\mu}}{\sqrt{4\pi T}}\, 
(\sigma_{\mu})^{\alpha \dot{\alpha}}\bar{S}_{\dot{\alpha}} \,\bar{S} \right] (\pt_{-} Y + \psi_{\phi}\psi_Y)
\,e^{ip_{\mu}x^{\mu}  - \sqrt{2}\phi- i \frac{Y}{\sqrt{2}}}.
\label{fermionV1}
\eeqn
In order to generate new bosonic vertex operators with the same mass \eqref{mass1} we apply supercharges to the fermion vertices above. Supercharge $Q$ acting on \eqref{fermionVdot1} gives the following bosonic vertices 
in the picture $(-1)$:
\beqn
&& \langle Q^{\alpha}, \bar{V}^{\dot{\alpha}}\rangle \sim \sigma_{\mu}^{\alpha\dot{\alpha}} 
\left( \psi^{\mu} + \frac{p^{\mu}}{\sqrt{4\pi T}}\, \psi\right) 
e^{-\varphi}\, e^{ip_{\mu}x^{\mu}  - \sqrt{2}\phi+ i \sqrt{2} \,Y}
\nonumber\\[3mm]
&&
= \sigma_{\mu}^{\alpha\dot{\alpha}}\left( V^{\mu}_{j=-1,m=1} 
+ \frac{p^{\mu}}{\sqrt{4\pi T}} V_{j=-1,m= 1}^{(-1)}\right)
\eeqn
where $V_{j=-1,m= 1}^{(-1)}$ is the scalar vertex \eqref{scalarj=-1}, while 
\beq
V^{\mu}_{j=-1,m=1} = \psi^{\mu} e^{-\varphi}\, e^{ip_{\mu}x^{\mu}  - \sqrt{2}\phi+ i \sqrt{2} m Y}
\label{vector}
\eeq
is a new vector vertex operator with $m=1$. We recognize it as a left-moving part of the spin-2 vertex \eqref{graviton}.
As was mentioned above, we obtained it by applying the supercharges to the scalar vertex \eqref{scalarj=-1}.
In a similar way we can generate the complex-conjugated vector $V^{\mu}_{j=-1,m= -1}$ with $m=-1$ if we apply the supercharge $\bar{Q}$ to the fermion vertex \eqref{fermionV1}.

We can also apply the supercharge $\bar{Q}$ to the fermion vertex \eqref{fermionVdot1}. This gives 
\beq
\langle Q^{\dot{\alpha}}, \bar{V}_{\dot{\beta}}\rangle \sim \delta^{\dot{\alpha}}_{\dot{\beta}}\,V_{j=-1, m=0},
\eeq
where
\beq
V_{j=-1, m=0}^{(-1)} = \left( \bar{\psi} + \frac{p^{\mu}}{\sqrt{4\pi T}}\,\psi_{\mu}\right)\, \pt_{-} Y\,
e^{-\varphi}\, e^{ip_{\mu}x^{\mu}  - \sqrt{2}\phi}
\label{m=0scalar}
\eeq
is a new scalar vertex with $m=0$ and mass \eqref{mass1}. Similarly, the action of $Q$ on the fermion vertex \eqref{fermionV1} gives
a complex-conjugated scalar vertex with the replacement $\bar{\psi} \to \psi$.

Finally, instead of the scalar vertex \eqref{scalarj=-1} we can start from another scalar vertex,
\beq
\tilde{V}_{j=-1,m= 1}^{(-1)} = \bar{\psi}\,e^{-\varphi}\,
e^{ip_{\mu}x^{\mu} - \sqrt{2}\phi  + i\sqrt{2}m  Y}\,.
\label{tildescalarj=-1}
\eeq
Note that this vertex is different from the one complex-conjugated to \eqref{scalarj=-1} because here 
we take $m=1$.
Conjugated to \eqref{tildescalarj=-1} is obtained by replacement $\bar{\psi} \to \psi$ and taking $m=-1$.

Following the same steps as above in the case of the vertex \eqref{scalarj=-1} one can show
that the action of supercharges on the scalar vertex \eqref{tildescalarj=-1} produces the same states which we already obtained
 from \eqref{scalarj=-1}.

Summarizing, in the bosonic left-moving sector for  $j=-1$ multiplet we find a  complex vector vertex \eqref{vector} and three
 complex scalar vertices, 
\beq
V_{j=-1,m= \pm 1}^{(-1)}, \qquad  \tilde{V}_{j=-1,m= \pm 1}^{(-1)}, \qquad V_{j=-1, m=0}^{(-1)}\,,
\label{3scalars}
\eeq
given by \eqref{scalarj=-1}, \eqref{tildescalarj=-1} and \eqref{m=0scalar} respectively.

\subsection{Building spin-2 multiplet} 

Now we will  use bosonic and fermionic vertex operators from the previous subsection 
 to construct the supermultiplet with $j=-1$ and mass \eqref{mass1}. Let us start with the R-R sector.
In much the same way as for the excited state in Sec. 5.2 we arrive at 
\beqn
V_{\dot{\alpha}\alpha}^{j= -1} &=& \bar{V}^L_{\dot{\alpha}}\left(m=\frac12\right) \times V^R_{\alpha}\left(m=-\frac12\right), 
\nonumber\\[3mm]
\bar{V}_{\alpha\dot{\alpha}}^{j=-1} &=& V^L_{\alpha}\left(m=-\frac12\right) \times \bar{V}^R_{\dot{\alpha}}\left(m=\frac12\right)\, , 
\label{RR1}
\eeqn
where the fermion vertices are given by \eqref{fermionVdot1} and \eqref{fermionV1}.
Expanding $V_{\dot{\alpha}\alpha}^{j= -1}$ and $\bar{V}_{\alpha\dot{\alpha}}^{j=-1}$ as in \eqref{expansion3/2}
we get a complex vector and a complex 3-form. As was discussed in Sec. 5.2, the massive 3-form dualizes into
a massive scalar. Thus in the R-R sector we get one complex vector and one complex scalar.

Now we pass to the NS-NS sector. The scalar vertices \eqref{3scalars} give $3\times 3 =9$ scalars of the form
\beq
V_i^L(m \ge 0)\times V_j^R (m\le 0)\,,
\label{9scalars}
\eeq
where $V_i(m)$, $i=1,2,3$, are given by \eqref{scalarj=-1}, \eqref{tildescalarj=-1} and \eqref{m=0scalar}, respectively.
Changing the sign of $m$ together with the  replacement $\psi \to \bar{\psi}$ gives nine complex conjugated scalars 
in addition to those in \eqref{9scalars}.

Combining the vector vertex \eqref{vector} with three scalar vertices \eqref{3scalars} provides us with six vectors of the 
form
\beqn
&&(V^{\mu}_{j=-1,m =1})^{L}\times V_j^R (m\le 0)\,, \nonumber \\[2mm] &&V_i^L(m\ge 0)\times (V^{\mu}_{j=-1,m=-1})^{R}\,,
\nonumber\\[2mm]
&& i=1,2,3.
\label{6vectors}
\eeqn
Again changing the sign of $m$ together with the  replacement $\psi \to \bar{\psi}$ gives six complex conjugated vectors to those.

Finally we can combine two vector vertices \eqref{vector} to produce a tensor
\beq 
(V^{\mu}_{j=-1,m =1})^{L}\times (V^{\nu}_{j=-1,m=-1})^{R}\,.
\label{tensor}
\eeq
Changing the sign of $m$ gives a complex conjugated tensor.
In 4D a massive vector has $(D-1)= 3$ physical degrees of freedom. Therefore for the tensor state
\eqref{tensor} we get
\beq
3\times 3 =9 = 5+3+1 \Rightarrow  1_{{\rm spin-2}} +1_{{\rm vector}} + 1_{{\rm scalar}}
\label{tensorcontent}
\eeq
massive degrees of freedom, where we show the expansion of the massive tensor into irreducible representations of 
SO$(D-1=3)$. Thus, from the 
complex tensor \eqref{tensor} we obtain one  spin-2 state, one vector and one scalar,  all of them complex.

Combining all bosonic states together we get
\beq
1_{{\rm spin}-2} + 8_{{\rm vector}} + 11_{{\rm scalar}}\, ,
\label{altogether}
\eeq
where we show the numbers of states with the  given spin.

How they split into 4D \ntwo supermultiplets? Long \ntwo spin-2 multiplet contains \cite{Zinoviev}
\beq
({\mathcal N}=2\;)_{{\rm spin-2}} =  1_{{\rm spin}-2} + 6_{{\rm vector}} + 1_{{\rm scalar}}
\label{spin-2}
\eeq
bosonic spin states
while long \ntwo vector multiplet has 
\beq
({\mathcal N}=2\;)_{ {\rm vector}} =   1_{{\rm vector}} + 5_{{\rm scalar}}
\eeq
bosonic spin states, see  Appendix B and Eq. \eqref{N=2vector}.

We conclude that $j=-1$ states with mass \eqref{mass1} form
\beq
(j=-1)\, {\rm states} = 1\times ({\mathcal N}=2\;)_{{\rm spin-2}} + 2\times ({\mathcal N}=2\;)_{ {\rm vector}} 
\label{j=-1states}
\eeq
(one spin-2 and two vector) \ntwo long (non-BPS) supermultiplets, all complex.

\section{Regge trajectories}
\label{Regge}
\setcounter{equation}{0}

In this section we will show  that all states we discussed in this paper  (shown in Fig. \ref{fig_spectrum}) are the
lowest states of the corresponding linear 
Regge trajectories. To construct these Regge trajectories we multiply the vertex operators \eqref{tachyon} or
\eqref{graviton} by derivatives of flat 4D coordinates. For example, for the scalar vertices \eqref{tachyon} we construct 
a family of vertices 
\beq
 \prod_{i=1}^{n} \pt_{-} x^{\mu_i}\pt_{+}x^{\nu_i}\,e^{-\varphi_L -\varphi_R }\, e^{ip_{\mu}x^{\mu}}\,
 V^{S, L}_{j=-\frac12,m}\times V^{S,R}_{j=-\frac12,-m}\, ,
\label{Reggetachyon}
\eeq
where $n$ is  $n=0,1,2,...$. The hadronic states associated with these vertices have at most spin $2n$. Their   mass  is
\beq
\frac{\left(M_{j=-\frac12, m}\right)^2}{8\pi T} = m^2 -\frac14 +n\,, \qquad n=0,1,2,...
\label{Reggetachyonmass}
\eeq
We see that mass squared for  these states depends linearly on  the spin. This linear Regge dependence 
appears because  we use the flat 4D part of the string $\sigma$ model to construct the Regge trajectories.

A similar construction can be developed for vertices \eqref{graviton}. Masses of these states are 
\beq
\frac{\left(M_{j=- 1, m}\right)^2}{8\pi T} = m^2  +n\,, \qquad n=0,1,2,...
\label{Reggegravitonmass}
\eeq
We have the same linear dependence with the same slope.

\section{Conclusions}
\label{conclusions}

In \cite{SYcstring} we observed that a vortex string supported in ${\cal N}=2$ SQCD is critical provided the following conditions are met:

(i) The gauge group of the model considered is U$(2)$;

(ii) The number of flavor hypermultiplets is $N_f=2N =4$;


The 4D theory under consideration is not conformal because of the Fayet-Iliopoulos parameter $\xi\neq 0$.
However, the gauge coupling  $\beta$ function vanishes; the Fayet-Iliopoulos parameter does not run either.

In addition to four translational zero modes this string exhibits three orientational and three size zero modes. Their geometry is described
by a non-compact six-dimensional Calabi-Yau manifold, the so-called resolved conifold $Y_6$.  The target space takes the form
$\mathbb{R}^4\times Y_6$. The emergence of six extra zero modes on the string under consideration makes the target-space model conformal,
the overall Virasoro central charge (including the ghost contribution) vanishes. Thus, this string is critical. The phenomenon we observed could be 
called a ``reverse holography."

The next question which was natural to address was the quantization of this closed critical string and the derivation of the
hadronic spectrum. The present paper completes the work started in \cite{KSYcstring,KSYconifold,SYlittles}. We calculated the masses of the massive
spin-0 and spin-2 states and constructed the 4D supermultiplets to which they belong. Our formulas match the previous result for the massless states.

The massive supermultiplets are shown to be long (non-BPS saturated).
We also prove that the above states are the lowest states on the corresponding Regge trajectories which are linear and parallel.

\section*{Acknowledgments}

The authors are grateful to Edwin Ireson  for useful  discussions and to Efrat Gerchkovitz and Avner Karasik
for helpful communications.
This work  is supported in part by DOE grant DE-SC0011842. 
The work of A.Y. was  supported by William I. Fine Theoretical Physics Institute  at the  University 
of Minnesota and  by Russian Foundation for Basic Research Grant No. 18-02-00048. 

\newpage 

\section*{Appendix A. BRST operator and vertices in the picture (0)}

 \renewcommand{\theequation}{A.\arabic{equation}}
\setcounter{equation}{0}

To convert vertex operator in the picture $(-1)$ into picture $(0)$ we use the BRST operator as follows
\cite{FriedMartShenk}:
\beq
V^{(0)} = \left\langle Q_{BRST}, \zeta \, V^{(-1)} \right\rangle,
\label{convert}
\eeq
where 
\beq
Q_{BRST} = \frac{1}{2\pi i}\, \int dz\left[ c T^m + \gamma G^m + \frac12 (c T^{gh} + \gamma G^{gh}) \right].
\label{BRST}
\eeq
Here $c$ and $\gamma$ are the ghosts of fermionic $(b,c)$ and bosonic $(\beta, \gamma)$ systems, respectively, 
while $T^m$, $G^m$   and $T^{gh}$, $G^{gh}$ are the energy momentum tensor and the supercurrent for matter and ghosts.
Below we will need the explicit expression for the matter supercurrent,
\beq
G^m = i\left( \psi^{\mu}\pt_{-} x_{\mu} + \psi_{\phi}\pt_{-}\phi + \psi_Y \pt_{-} Y\right).
\label{Gm}
\eeq
The ghost system $(\beta, \gamma)$ can be expressed in terms of fermions $\eta$, $\zeta$,
\beq
\gamma = e^{\varphi}\eta, \qquad \beta = e^{-\varphi}\pt_{-} \zeta, 
\eeq
where the propagator of $\eta$, $\zeta$ is normalized as
\beq
\langle \eta (z), \zeta(0)\rangle = \frac1z\,.
\eeq

To convert the left-moving part of the scalar vertex \eqref{tachyon} in the picture $(-1)$ into the picture $(0)$ we use
the rule \eqref{convert}. We arrive at the expression \eqref{tachyon0} with the help of \eqref{Gm}.

Similarly, for the  $j=-1$ vertex \eqref{scalarj=-1} we again use \eqref{Gm} to obtain the vertex operator 
\eqref{picture0j=-1} in the picture (0).

\newpage

\section*{Appendix B. Long \boldmath{\ntwo} vector and spin-2 \\  multiplets in 4D.}

 \renewcommand{\theequation}{B.\arabic{equation}}
\setcounter{equation}{0}

In this Appendix we briefly review construction of \ntwo long massive supermultiplets in four dimensions.
For massive states in the rest frame supersymmetry generators $Q^{\alpha f}$ and $\bar{Q}_{f \dot{\alpha}}$ can be viewed as annihilation
and creation operators, where $f=1,2$ is the index of two ${\cal N}=1$ supersymmetries which constitute \ntwo\!. Assuming that
the annihilation  operators $Q^{\alpha f}$ produce zero upon acting on a ``ground state''  $|a\rangle $ we can generate 
all states of a given supermultiplet applying to $|a\rangle $ the creation operators $\bar{Q}_{f \dot{\alpha}}$.

For simplicity we will consider only the bosonic states in  a multiplet. Assuming that $|a\rangle $ is a bosonic state we have
6 possibilities 
\beq 
\left\{ \bar{Q}_{1 \dot{1}} \bar{Q}_{2 \dot{1}}, \; \bar{Q}_{1 \dot{1}} \bar{Q}_{1 \dot{2}}, \;
\bar{Q}_{1 \dot{1}} \bar{Q}_{2 \dot{2}}, \; \bar{Q}_{2 \dot{1}} \bar{Q}_{1 \dot{2}}, \;
\bar{Q}_{2 \dot{1}} \bar{Q}_{2 \dot{2}}, \; \bar{Q}_{1 \dot{2}} \bar{Q}_{2 \dot{2}} \right\} \times |a\rangle
\label{level2}
\eeq
at level 2 and only one possibility 
\beq
\bar{Q}_{1 \dot{1}} \bar{Q}_{2 \dot{1}}\bar{Q}_{1 \dot{2}} \bar{Q}_{2 \dot{2}}\times |a\rangle
\label{level4}
\eeq
at level 4.

First let us construct the long \ntwo massive vector supermultiplet. In this case we choose $|a\rangle$ to be a scalar with spin $J=0$. The construction is shown in Table~\ref{table1} where $J_z$ is the $z$-projection of spin
and level $0$ denotes the state $|a\rangle$ itself.
\begin{table}
\begin{center}
\begin{tabular}{|c|c | c| c|c|}
\hline
$\rule{0mm}{6mm}$ $J_z$ & Level $0$ & Level 2 & Level 4 & Sum 
\\[3mm]
\hline
$\rule{0mm}{5mm}$ $1$ & $0$ & 1 & $0$ & 1 
\\[3mm]
\hline
$\rule{0mm}{5mm}$ $0$ & 1   & 4 & 1   & 6 
\\[3mm]
\hline
$\rule{0mm}{5mm}$ $-1$ & $0$ & 1 & $0$ & 1 
\\[3mm]
\hline
\end{tabular}
\end{center}
\caption{{\small  Structure of the vector multiplet.
We show the numbers of states with the given $J_z$ produced by the action of supercharges at each level and their sum.}}
\label{table1}
\end{table}
Here we used the fact that, say,  in Eq. \eqref{level2} the product $\bar{Q}_{1 \dot{1}} \bar{Q}_{2 \dot{1}}$ acting on $|a\rangle $
increases $J_z$ by one, 
four product operators of the type $\bar{Q}_{ f\dot{1}} \bar{Q}_{g \dot{2}}$ $(f,g=1,2)$ do not change $J_z$, while the product
$\bar{Q}_{ 1\dot{2}} \bar{Q}_{2 \dot{2}}$ reduces $J_z$ by one.

Overall we observe one state with $J_z =1$, one state with $J_z = -1$ and 6 states with
$J_z=0$. This gives the decomposition \eqref{N=2vector}.

Now let us pass to the spin-2 supermultiplet. To this end we take  $|a\rangle$ to be a vector state
with spin $J=1$. The resulting structure is shown in Table~\ref{table2}.
\begin{table}
\begin{center}
\begin{tabular}{|c|c | c| c|c|}
\hline
$\rule{0mm}{6mm}$ $J_z$ & Level $0$ & Level 2 & Level 4 & Sum 
\\[3mm]
\hline
$\rule{0mm}{5mm}$ $2$ & $0$ & 1 & $0$ & 1 
\\[3mm]
\hline
$\rule{0mm}{5mm}$ $1$ & $1$ & 4+1=5 & $1$ & 7
\\[3mm]
\hline
$\rule{0mm}{5mm}$ $0$ & 1   & 4+1+1=6 & 1   & 8 
\\[3mm]
\hline
$\rule{0mm}{5mm}$ $-1$ & $1$ & 4+1=5 & $1$ & 7 
\\[3mm]
\hline
$\rule{0mm}{5mm}$ $-2$ & $0$ & 1 & $0$ & 1 
\\[3mm]
\hline
\end{tabular}
\end{center}
\caption{{\small  Spin-2 multiplet.
}}
\label{table2}
\end{table}
Here, say, 4+1+1=6 means that at level 2 four states are generated from the state at level 0 in the same row,
while two other states come from states at level 0 in the upper or the lower neighboring rows. The last column gives
decomposition \eqref{spin-2}.

\end{document}